# Tracking atomic-scale interdiffusion in immiscible bimetallic nanoparticles via four-dimensional electron tomography

**Authors:** Jisheng Xie, Dijin Jiang, Zhen Sun, Yiheng Dai, Zezhou Li, Yao Zhang, Jihan Zhou*

**Affiliations:**

Beijing National Laboratory for Molecular Sciences, Center for Integrated Spectroscopy, College of Chemistry and Molecular Engineering, Peking University; Beijing, 100871, China.

*Corresponding author. Jihan Zhou, jhzhou@pku.edu.cn

**Abstract:** The mixing behavior of multi-element nanomaterials often diverges from bulk equilibrium, yet a quantitative and atomically-resolved description of this transformation has remained challenging. Here, using ex situ four-dimensional atomic-resolution electron tomography combined with in situ scanning transmission electron microscopy, we reveal the atomic-scale pathway of the miscible transition in nominally immiscible core-shell Pd@Ir nanoparticles. The transformation begins with surface reconstruction via atom hopping (200 °C) followed by surface flattening (300 °C). The critical transition occurs at 400 °C, where Ir interfacial diffusion (diffusion coefficient = $8.0 \times 10^{-24}$ $m^2$ $s^{-1}$) initiates, accompanied by the emergence of discrete rather than collective Ir centers as intermediates that drives the system into a miscible arrangement. Upon reaching the nanoscale melting point (900 °C), Ir atoms collectively diffuse inward, ultimately yielding the thermodynamically stable Ir@Pd core-shell configuration. Our work provides quantitative atomic-scale insights into how a metastable nanostructure evolves through distinct intermediates towards equilibrium, offering a framework for the design of advanced multi-element materials.

**Main Text:** Multielement nanomaterials lie at the forefront of materials design, offering unparalleled tunability in catalytic (*1, 2*), energy storage (*3, 4*), and electronic properties (*5, 6*) through synergistic interactions between constituent elements. Traditionally, their phase stability and transformation pathways are predicted from bulk equilibrium phase diagrams. However, at the nanoscale, high surface-to-volume ratios and spatial confinement can drive systems far from bulk thermodynamics (*7, 8*), enabling metastable structures and novel functionalities (*9–12*). A particularly profound yet poorly understood phenomenon is the atomic-scale interdiffusion of elements that are immiscible in the bulk. Quantifying how and when these nominally immiscible atoms diffuse across interfaces at the nanoscale is critical, as it dictates the ultimate stability, compositional distribution, and performance of the material. Unraveling the dynamic atomic pathway in four dimensions (space plus time)—beyond static snapshots—represents a fundamental step toward rationally designing next-generation multielement materials with tailored properties and lifetime. Yet, direct, quantitative observation of such diffusion processes with four-dimensional atomic resolution remains a formidable challenge in materials science. While aberration-corrected scanning transmission electron microscopy (STEM) provides atomic-resolution imaging (*13, 14*) and has enabled studies of diffusion (*15, 16*) and phase transitions (*17–19*), conventional two-dimensional (2D) projections often lack the sensitivity to detect subtle, early-stage atomic intermixing. Atomic-resolution electron tomography (AET) overcomes this limitation by resolving three-dimensional (3D) atomic positions (*20–24*) and chemical identities (*25–30*). Extending this approach, four-dimensional AET (4D-AET) allows continuous tracking of atomic-scale dynamics under controlled conditions (*31, 32*), making it uniquely suited to quantify the onset and pathway of diffusion in immiscible systems. Here, using 4D-AET (treated 200 - 400°C ) combined with in situ STEM (from room temperature to 900 °C), we investigate the atomic-scale evolution of initially immiscible Pd@Ir core-shell nanoparticles (NPs) across a temperature range from 200 °C to the nanoscale melting point ($T_m$ ~900 °C, Fig. 1A and movie S1). We identify a critical transition temperature ($T_c$ ~400 °C), below which surface reconstruction proceeds via atom hopping and surface flattening. At $T_c$, Ir interfacial diffusion initiates with a measured coefficient of $8.0 \times 10^{-24}$ $m^2$ $s^{-1}$, mediated by discrete Ir intermediates that drive the system toward a miscible state. This reduces the miscibility gap from 0.81 $T_m$ in bulk to 0.57 $T_m$ at the nanoscale. Ultimately, near the melting point, collective inward diffusion of Ir leads to the thermodynamically stable Ir@Pd configuration, revealing a clear, quantifiable pathway from metastability to equilibrium (Fig. 1A).

## Atomic-scale evolution of the miscible transition

As a model system, we selected Pd and Ir—elements that are nearly immiscible in bulk (solubility <4 wt%; fig. S1) (*33*) below 500 °C but share similar metallic radii and air stability, facilitating unambiguous structural and chemical analysis during transformation. We synthesized Pd@Ir core–shell NPs (6-7 nm in diameter) with the face-centered cubic (FCC) symmetry (*34*) and characterized them using annular dark-field STEM (ADF-STEM) and energy-dispersive X-ray spectroscopy (EDS), confirming the core–shell structure prior to heating (See materials and methods and fig. S2). Because Pd exhibits a lower surface energy (2.05 J $m^{-2}$) than Ir (3.00 J $m^{-2}$) (*35*), Pd is thermodynamically favored to occupy surface sites upon annealing, driving the initial Pd@Ir architecture toward a more stable configuration.

To track the thermally induced structural evolution of Pd@Ir NPs at nanoscale, we performed STEM using an in situ heating holder across a temperature range (See materials and methods and fig. S3). At room temperature, the Pd@Ir NPs exhibit a cubic core-shell morphology (Fig. 1A and fig. S3). Owing to the inherent limitation of 2D imaging, pronounced changes in image contrast only emerge at around 800 °C, where the core–shell interface becomes visibly blurred, indicating that the thermal energy is now sufficient to drive rapid Pd-Ir intermixing (fig. S3). Upon further heating to 900 °C, the particles progressively round, and the Pd core exhibits enhanced mobility. Concurrently, Ir atoms collectively diffuse inward, fluctuate within the mobile Pd matrix, and eventually consolidate into a distinct Ir core, yielding the thermodynamically stable Ir@Pd configuration (movie S1). These observations confirm that the NPs reach their nanoscale $T_m$ (~900°C), which is substantially lower than the bulk $T_m$ of 1300°C (fig. S1), and complete the transition to the equilibrium structure.

To resolve the atomic diffusion during the initial miscible transition (200-400°C), we performed 4D-AET experiments (see materials and methods). Tomographic tilt series were acquired for three individual NPs (Particles 1-3; twelve datasets in total) using the ADF-STEM mode, each annealed at different temperatures (figs. S4 to S15). Particles 1-3 at different annealing stages are designated as $Pd@Ir_{m_n}$, m denotes the annealing temperature (200, 300, or 400 °C), and n denotes the total annealing time (0, 30, 60, or 120 min). All data were collected at reduced electron dose (tables S1-3) compared to prior work (*31*) (figs. S16 to S18). Representative images reveal increasingly pronounced shape changes as the temperatures rises from 200 to 400°C (Fig. 1B). Through image preprocessing, 3D reconstruction, atom tracing, and chemical classification, we obtained full experimental 3D atomic models for each particle at every annealing stage (see materials and methods, Fig. 1, C to F and fig. S19). The pristine Pd@Ir particles exhibit a cuboctahedral morphology with element-mixed surfaces in the Ir shells (Fig. 1, C to E), consistent with our recent observations (*29, 36*), and possess an averaged Pd weight percent of ~ 69% (tables S1-3). Using the experimentally resolved atomic models, we performed a quantitative structural analysis (see materials and methods and tables S1-3). At 200 and 300°C, the total atom numbers and surface areas of Particles 1 and 2 remained nearly constant (fig. S20A). For Particle-1 at 200°C, atoms in the Pd-Ir mixing surface exhibited thermal fluctuation after successive annealing, with no obvious elemental segregation (Fig. 1C and fig. S19, A to D). At 300°C, partial Pd segregated onto the surface of Particle-2 (Fig. 1D and fig. S19, E to H). At 400 °C, Particle-3 showed a slightly increase in both atom number and surface area (fig. 20A), attributable to the Oswald ripening (*37, 38*). Concurrently, more pronounced Pd segregation occurred on its surface (Fig. 1E and fig. S19, I to L). By comparing the contributions from Oswald ripening and Pd segregation to surface composition (fig. S20B), we found that the Pd-enriched surface at this temperature is primarily driven by Pd segregation. By examining central slices from each atomic model (fig. S21), we further revealed that although most particles retained a core-shell morphology, Ir atoms began to diffuse inward after 120 min of annealing at 400 °C, marking the onset of the miscible transition at this critical temperature.

Our observations demonstrate that annealing between 200 and 400 °C progressively enhances surface reconstruction, which is accompanied by Pd segregation and the onset of Ir inward diffusion. The miscible transition is found to initiate at 400°C.

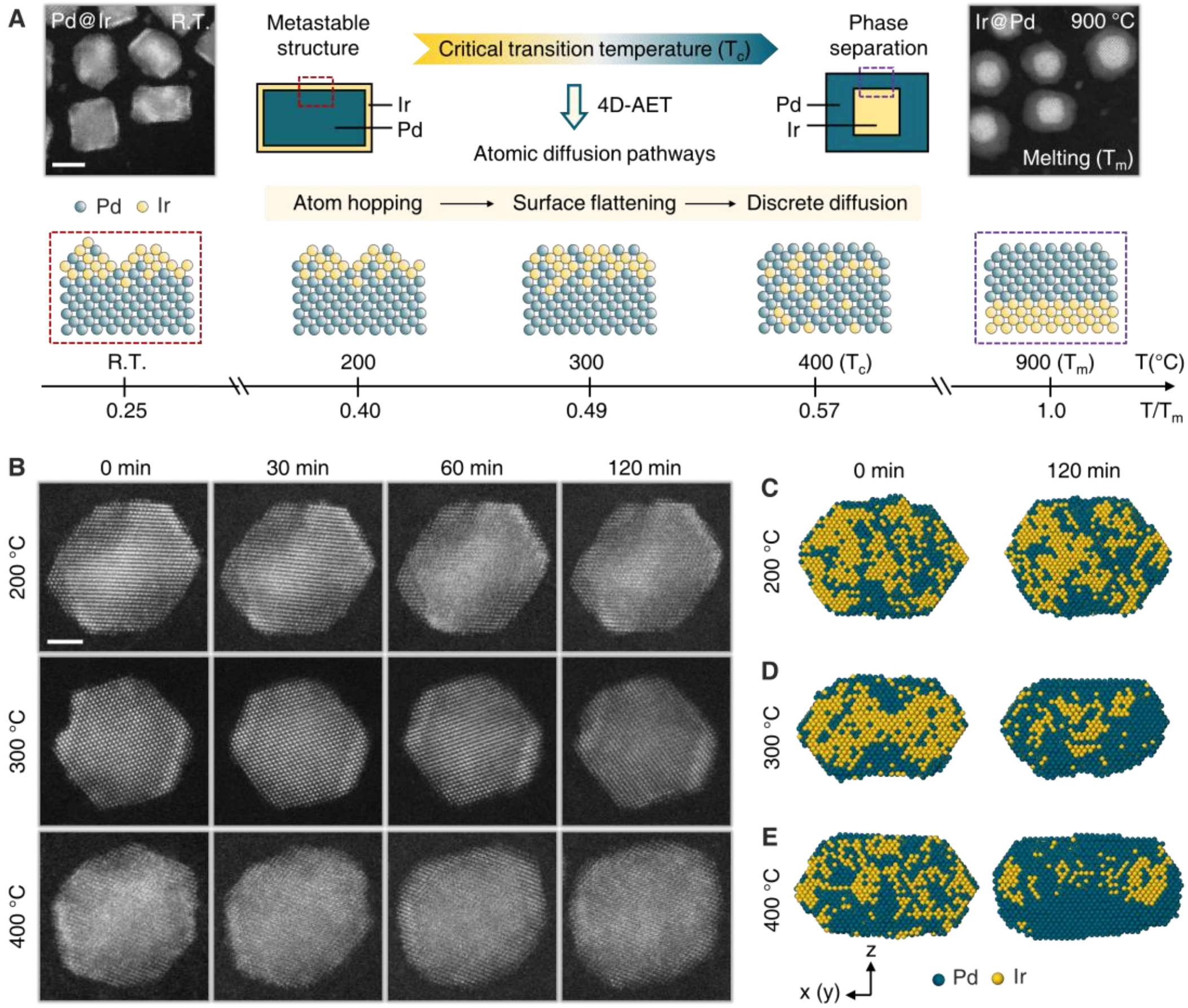

**Fig. 1. Overview of the miscibility features in Pd@Ir.** (**A**) In situ ADF-STEM images of Pd-Ir NPs at R.T. and 900°C, and three schematic pathways in the miscible transition between 200 and 400°C revealed by 4D-AET. The Pd@Ir NPs undergo "atom hopping, surface flattening, and discrete diffusion" and ultimately transitions into the thermodynamically stable Ir@Pd configuration at $T_m$. At the nanoscale, $T_c/T_m$ is reduced to 0.57. (**B**) Representative ADF-STEM images of Particles 1-3 at different annealing stages. (**C** to **E**) Experimental 3D atomic models of Particles 1-3 after 0-min and 120-min annealing at 200°C (C), 300°C (D), and 400°C (E), respectively. The full time series of 3D atomic models at 0, 30, 60 and 120 min is shown in fig. S19. Green and yellow represent Pd and Ir atoms, respectively. Scale bars: 5 nm for images in (A) and 2 nm for images in (B).

## Quantifying atomic pathways in surface reconstruction and Pd segregation

To elucidate the atomic diffusion pathways leading to the miscible transition, we quantified the 3D atomic evolution of atomic positions, mapped the local migration pathways during surface reconstruction, and evaluated their impact on the chemical order in near-surface regions (see materials and methods).

First, by comparing the 3D atomic models before and after specific annealing intervals, we identified the displaced or chemically altered atoms (inconsistent atoms, unpaired atomic positions or changed chemical species) to define the spatial extent of migrated atoms (see materials and methods and figs. S22-25 for detailed analysis of unpaired atomic positions). From 200 to 400°C, the number of migrated atoms progressively increased, with their migration depths extending mainly from 8 Å to 12 Å below the surface (Fig. 2A). At 400 °C, a small number of migrated atoms were observed beyond 12 Å; these correspond to the onset of Ir inward diffusion (fig. S21). Collectively, the migration profiles indicate a gradual activation of interior atoms as the annealing temperature rises. We also evaluated the chemical consistency across annealing stages. At 200 °C, the atomic models showed ~92% chemical consistency (table S1), which lies close to the 95–97% structural consistency, a reproducibility limit established in prior AET studies of metallic nanoparticles (*31, 39*). The high consistency implies minimal structural evolution at this stage. At elevated temperatures, atomic diffusion becomes more pronounced, reducing the consistent atom proportion to ~90% at 300 °C and to ~84% at 400 °C (tables S2 and S3).

Next, to resolve the local atomic pathways of surface reconstruction, we tracked structural changes within an island-like region (enriched in low-coordinated atoms) in the same particle across successive annealing steps. At 200 °C, the dominant diffusion mechanism is "atomic hopping" between adjacent sites in Particle-1, which largely preserved the convex morphology of the island region throughout 120 min of annealing (Fig. 2B). At 300 °C, surface reconstruction intensified: the island-like region of Particle-2 was flattened via atomic migration, losing its convex features within 30 min and maintained this smoother morphology thereafter (Fig. 2C). By 400 °C, the island-like region of Particle-3 similarly vanished after 30 min (Fig. 2D). In later annealing stages, a slight increase in the number of atomic layers occurred due to continued surface reconstruction (fig. S20A). These local migration trends are consistent with the evolution of the overall surface curvature: as temperature rises, the absolute curvature values decrease, reflecting the progressively flattening of convex and concave surface features (fig. S26).

Then, we further quantified the redistribution of Pd and Ir near the surface of Particles 1-3 by calculating the surface Pd concentrations and the chemical order (characterized by chemical short-range-order parameters, CSROPs) for each annealing step (see materials and methods). At 200 °C, where atomic hopping dominates, neither the surface Pd concentration (50 ± 4 %) nor the chemical order changed significantly (Fig. 2E and fig. S27). 3D rendering (yellow regions in fig. S27, A to D) and histogram analysis (negative-value parts of the histogram in fig. S27E) confirm that the chemical order in surface and subsurface regions of Particle-1 changed little. At 300 °C, 30 min of annealing induced the surface flattening (Fig. 2B) and altered compositions: the surface Pd concentration increased by ~13 % (Fig. 2E); Ir-rich regions contracted (minimum CSROP increased from –4 to –3). and the Pd-rich regions expanded (proportion of CSROP near +1 increased) (fig. S28E). The structure then stabilized during further annealing, with only minor fluctuations in surface Pd concentration (variations ≤0.3 %) and chemical order. At 400 °C, more pronounced structural evolution drove a continuous increase in surface Pd concentration from 58 % to 78 % (Fig. 2E). Pd enrichment was also clearly reflected in the chemical order after each annealing step (Fig. 2, F to J and fig. S29).

To determine whether the evolution of chemical order depends on crystallographic orientation, we divided each particle into eight {111}-oriented and six {100}-oriented regions based on its cuboctahedral symmetry (see materials and methods and fig. S30). By calculating the difference in averaged CSROPs between successive annealing stages (ΔCSROPs), we tracked the overall variation in elemental enrichment within each region (Fig. 2, J to L). At 200 °C, all ΔCSROPs were similar across orientations, varying within ±0.2 (Fig. 2J), consistent with the limited extent of surface reconstruction observed at this temperature. At 300 °C, pronounced changes in ΔCSROPs emerged in some {100} regions, whereas variations in {111} regions remained relatively small (Fig. 2K). The surface flattening at this stage triggered noticeable reshaping of Particle-2, including a reduction in thickness along the z-direction and elongation in the x-y plane (dashed lines in fig. S28). These shape changes, which are closely linked to atomic migration occurring preferentially along {100} orientation (fig. S23A), contributed to the evolution of ΔCSROPs in {100} regions. At 400 °C, although Oswald ripening altered the total atom number, the intensified structural changes including Pd segregation and Oswald ripening drove a clear increase in ΔCSROPs across all crystallographic orientations (Fig. 2L).

Our atomic-scale observations reveal a temperature-dependent evolution of surface reconstruction: it progresses from atomic hopping at 200 °C to surface flattening at 300 °C, resulting in pronounced Pd segregation at 400 °C. We quantify the consequent enrichment of Pd in near-surface regions and establish a direct correlation between the orientation-dependent chemical evolution and the macroscopic reshaping of the nanoparticles.

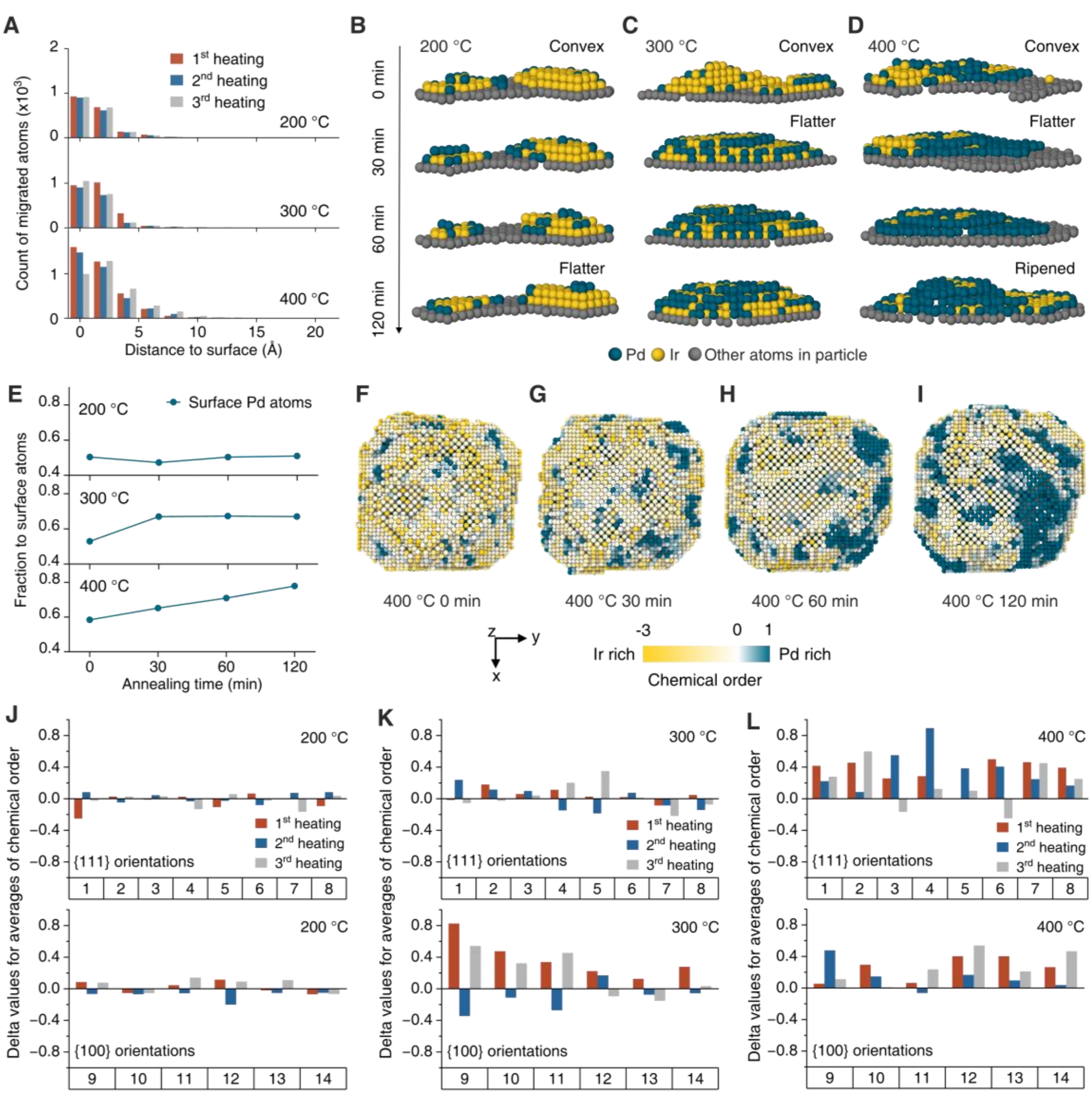

**Fig. 2. Quantification of 4D atomic evolution and chemical order.** (**A**) Distribution of distances to the surface for migrated atoms in Particles 1-3 at different annealing stages. (**B** to **D**) Surface reconstruction pathways within island-like regions at 200°C (atom hopping) (B), 300°C (surface flattening) (C) and 400°C (surface flattening) (D). Green, yellow and gray represent Pd, Ir and other atoms in particle, respectively. (**E**) Evolution of surface Pd concentration in Particles 1-3 at different annealing stages. Surface atoms are defined as atoms with coordination number<12. (**F** to **I**) 3D renderings of chemical order in Particle-3 after annealing for 0-min (F), 30-min (G), 60-min (H) and 120-min (I), respectively. The green, yellow and white colormap represents the Pd-rich regions, the Ir-rich regions and regions where the local composition equals the averaged Pd concentration in the whole particle, respectively. The color scale is normalized from -3 to 1. Histograms of CSROPs for Particles 1-3 are shown in figs. S27 to S29. (**J** to **L**) Histograms of ΔCSROPs for Particle-1 (J), Particle-2 (K) and Particle-3 (L). Surface atoms are grouped into eight {111}-oriented and six {100}-oriented regions, and the CSROPs are averaged within each region (fig. S30). ΔCSROPs in the {111} (upper panels) and {100} (lower panels) orientations are calculated between successive datasets before and after the first, second and third annealing steps. Each group of histograms is indexed according to the corresponding crystallographic planes among the 14 orientations.

## Ir discrete diffusion during the onset of the miscible transition

Discrete inward diffusion of Ir atoms occurs within the NP at 400°C (fig. S21). We further quantify the extent of this Ir inward diffusion across annealing stages and evaluate its contribution to the Pd–Ir miscible transition.

First, we mapped the 3D positions of interior Ir atoms (Fig. 3A) and calculated their distances to the particle surface at each annealing stage (Fig. 3B). At 200 °C, Ir atoms in Particle-1 are confined to within 10 Å of the surface, indicating no substantial inward diffusion. At 300 °C, the first annealing period drives a portion of interior Ir atoms in Particle-2 to depths beyond 10 Å; subsequent annealing at this temperature leaves the Ir distributions largely unchanged. By 400 °C, the number of interior Ir atoms increases progressively with annealing time. After 120 min, Ir atoms ultimately penetrate to regions approximately 20 Å from the surface in Particle-3.

Next, to quantify the degree of Pd–Ir mixing inside the NPs, we computed the layer-averaged chemical-order profiles and monitored the local chemical concentration ($C_{Ir}$ and $C_{Pd}$) around each fully coordinated bulk atomic center (coordination number (CN) = 12). The averaged chemical-order profiles for all pristine particles exhibit typical core-shell structures (0 min in Fig. 3C): an Ir-rich shell (blue), a Pd-rich core (red), and a mixed Pd–Ir interface (lighter color, near-white region). At 200 °C, the profiles remain nearly unchanged throughout annealing. At 300 °C, the interface (4-10 Å subsurface region) in Particle-2 becomes blurred after the first annealing period (0 to 30 min in Fig. 3C), a change that persists during subsequent annealing. This blurring correlates with particle reshaping (fig. S28) and rearrangement of near-surface atoms rather than with the onset of bulk intermixing. At 400 °C, the Pd–Ir interface progressively blurs after 30 min and 60 min of annealing. After 120 min, all atomic layers in Particle-3 exhibit lighter coloring compared to the pristine state, indicating the core–shell contrast is lost. These profile changes reflect a continuous tuning of the local chemical environment with increasing temperature and time.

Analysis of the local Ir concentration ($C_{Ir}$) around bulk Ir center (Fig. 3D) reveals that at 200 °C the $C_{Ir}$ distributions nearly overlap, peaking at ~60%. At 300 °C, the particle reshaping increases the numbers of bulk Ir centers without shifting the $C_{Ir}$ distribution, indicating that more Ir atoms have moved inward while remaining in Ir-rich local environments. At 400 °C, prolonged annealing further increases the numbers of bulk Ir centers. After 120 min, a noticeable rise in population of Ir-rich centers at $C_{Ir}$ = 0 and 8.3% (corresponding to only 1–2 Ir atoms in the coordination shell, Fig. 3D), confirms that a subset of Ir atoms have entered Pd-enriched regions via discrete Ir diffusion (through single- or dual-atom pathway, as marked by the solid arrow in Fig. 3D). These trends are further supported by the corresponding distributions of local Pd concentration ($C_{Pd}$) around bulk Pd centers (fig. S31).

To trace the local transition pathways, we compared identical regions within the atomic models of Particle-3 before and after 120 min of annealing at 400°C, where Ir atoms had diffused inward (Fig. 3E). To map the overall spatial distribution, we identified all discrete Ir centers together with their coordinated atoms (discrete Ir centers are defined as local Ir environments with $C_{Ir} = 0$ and 8.3%), and calculated their distances to the surface. In the pristine Particle-3 (left panel in Fig. 3F), only a small number of discrete Ir centers are presented, located predominantly near the surface (Fig. 3G). After 120 min of annealing, more discrete Ir centers emerge (right panel in Fig. 3F) and penetrate to depths of approximately 20 Å from the surface (Fig. 3G).

In the bulk system, Pd and Ir are phase-separated (fig. S1), where heteroatomic (Pd–Ir) bonds are energetically less favorable than the average of the homoatomic (Pd–Pd and Ir–Ir) bonds (*33*). Based on surface energy and bond energy, it has been calculated that Pd-Ir configurations with varied Ir contents reside at the higher energy than the phase-separated Ir@Pd state (*40*). Consequently, our results experimentally demonstrate that, starting from the metastable Pd@Ir architecture, the system evolves through an intermediate state mediated by discrete Ir diffusion, which triggers the onset of the miscible transition at the critical temperature $T_c$ of 400°C.

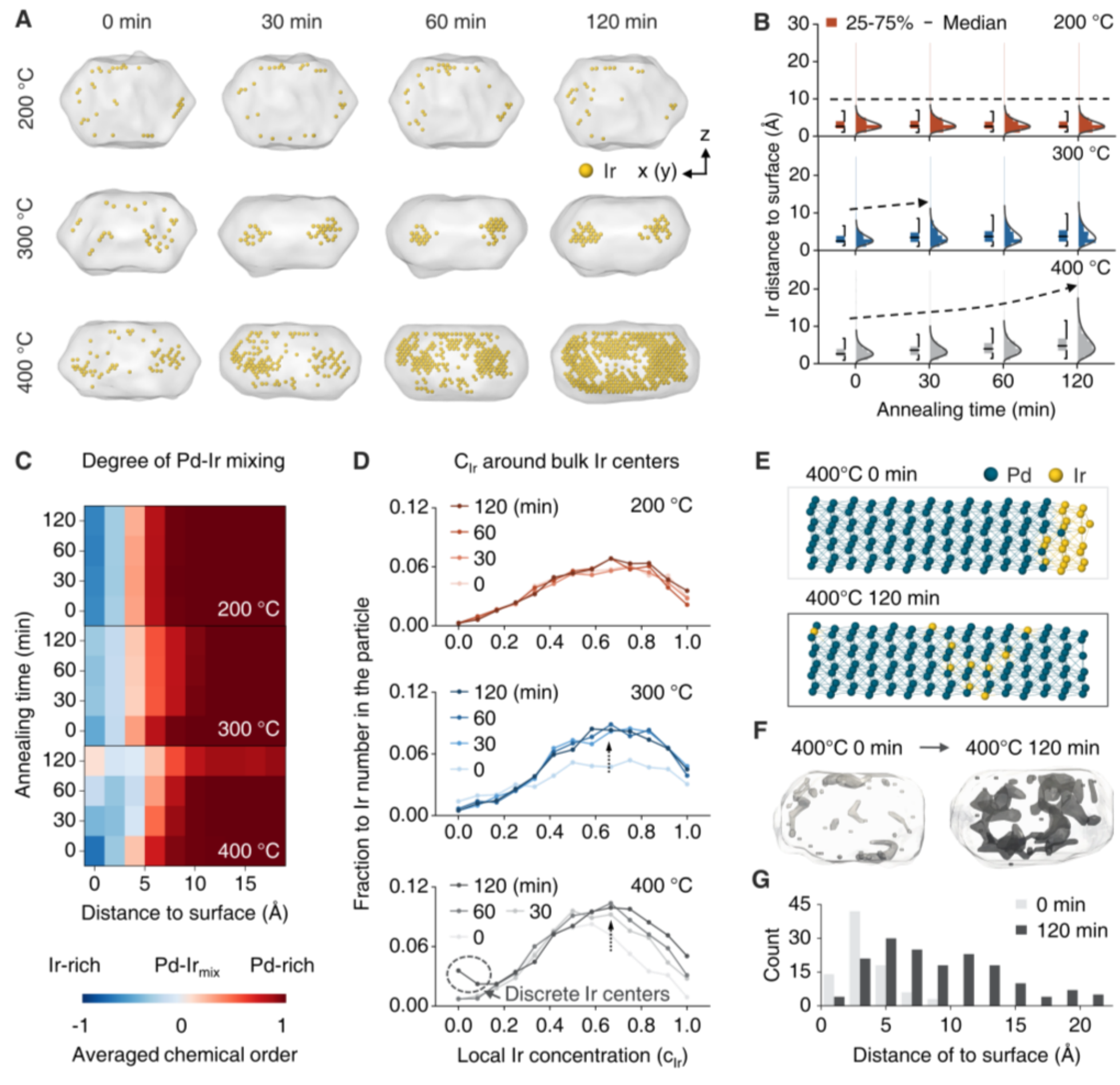

**Fig. 3. Ir discrete diffusion during the miscible transition.** (**A**) Interior Ir atoms in Particles 1-3 at different annealing stages. Outer Ir shellss are hidden to better visualize the interior Ir distribution. Gray meshes represent the overall nanoparticle shapes. (**B**) Distribution of all Ir atoms as a function of their distance to the surface. Dashed lines indicate the evolving trends across different annealing stages. (**C**) Layer-averaged chemical-order profiles for Particles 1-3. The blue, red and white colormap represent Ir-rich regions, Pd-rich regions and regions where the local composition equals to the averaged Pd concentration of the whole particle, respectively. (**D**) The distribution of local Ir concentrations ($C_{Ir}$) around bulk Ir centers. Bulk centers are defined as atoms with CN=12. $C_{Ir}$ (x axis) represents the Ir atomic fractions, Ir/(Pd+Ir), within the first coordination shells, indicating the local probability of finding Ir around a central atom. Dashed arrows highlight the increase in the population of bulk Ir-rich centers. The solid arrow marks the appearance of discrete Ir centers, characterized by $C_{Ir}$ = 0 and 8.3% (corresponding to only 1–2 Ir atoms in the coordination shell). (**E**) 3D renderings of identical regions in Particle-3 at 0-min and 120-min annealing. (**F**) 3D renderings of local environments with $C_{Ir}$ = 0 and 8.3%. The lightest-gray mesh outlines the shape of particle-3. Darker gray meshes enclose the regions with $C_{Ir}$ = 0 and 8.3% at 0-min (left) and 120-min (right) annealing. All discrete Ir centers and their coordinated atoms are displayed. (**G**) Distance-to-surface distribution for discrete Ir centers ($C_{Ir}$ = 0 and 8.3%) in Particle-3 after 0-min and 120-min annealing. Green and yellow in (A) and (E) represent Pd and Ir atoms, respectively.

### Miscible behavior at the nanoscale

At the nanoscale, discrete Ir diffusion emerges as key structural features in the early-stage of the miscible transition. To quantify the associated diffusion parameters, we monitored the evolution of interfacial Ir concentration within a 6 Å-thick subsurface shell (fig. S32 and Fig. 4A). Although at 300 °C, particle reshaping dominates the compositional reconstruction in the first annealing, the Ir concentration remains nearly constant at both 200 °C and 300 °C, indicating no significant interfacial diffusion. In contrast, at 400°C, the interfacial Ir concentration increases continuously from 3.2% to 24.3% over 120 min. Applying Fick's second law to this evolution (*41*), we calculate an interfacial diffusion coefficient of Ir atoms to be $8.0 \times 10^{-24}$ $m^2$ $s^{-1}$ at 400 °C (see materials and methods and Fig. 4B).

We summarize our key quantitative results in a schematic that compares the Pd–Ir bulk system (69 wt% in Pd) with the Pd@Ir NPs studied here (Fig. 4C). In the bulk Pd–Ir phase diagram, the miscibility gap extends up to approximately 0.81 $T_m$ (fig. S1). Our in situ heating experiments show that the $T_m$ drops from 1300 °C in the bulk to ~900 °C for the 6–7 nm Pd@Ir NPs. Furthermore, 4D-AET reveals that, driven by the higher surface energy of Ir relative to Pd, the miscibility gap of metastable Pd@Ir NPs is further reduced to approximately 0.57 $T_m$. This pronounced lowering of $T_c/T_m$ is even more substantial than the predicted $T_c$ for Pd–Ir clusters composed of hundreds of atoms (*42*). Upon reaching $T_c$, Pd@Ir NPs begin to transition from a heterostructured core-shell configuration into a miscible Pd–Ir solid solution, passing through an intermediate state characterized by discrete Ir centers. As the temperature approaches $T_m$, the available thermal energy becomes sufficient to drive rapid structural transformation, triggering collective inward diffusion of Ir atoms and ultimately yielding the thermodynamically stable Ir@Pd configuration.

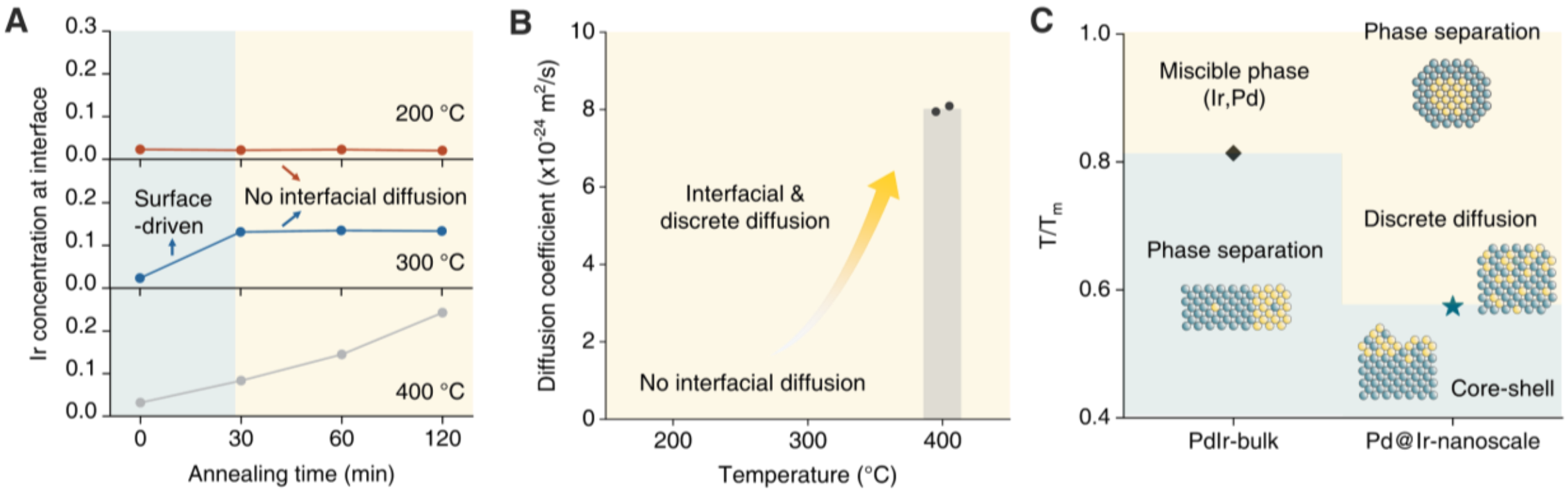

**Fig. 4. Measurement of interfacial diffusion coefficient and miscible behavior at the nanoscale.** (**A**) Radial Ir concentrations at the Pd-Ir interface within a 6 Å surface region. The Ir concentration at different depths are listed in fig. S32. At 300 °C, the surface-driven compositional reconstruction observed during the first annealing step is attributed to particle reshaping (fig. S28). Given that the Ir concentration remains nearly constant in subsequent annealing stages at 300 °C, and similar behavior is observed at 200 °C, we conclude that no significant interfacial diffusion was detected at either 200 °C or 300 °C in this system. (**B**) Measurement of the interfacial Ir diffusion coefficient. (**C**) Comparisons of miscibility behavior at the bulk and nanoscale.

## Summary and conclusions

In summary, we have used in situ atomic imaging and 4D-AET to quantify the atomic-scale evolution of the miscible transition in Pd@Ir NPs. Surface reconstruction below 400 °C advances through two distinct stages: atom hopping and surface flattening. The onset of miscibility at the critical temperature of 400 °C is marked by interfacial Ir diffusion (diffusion coefficient $8.0 \times 10^{-24}$ $m^2 s^{-1}$), which proceeds via discrete Ir atomic pathways. At nanoscale, surface energy suppresses the miscibility gap from 0.81 $T_m$ in the bulk to 0.57 $T_m$. These results reveal how a metastable core-shell nanostructure evolves through an intermediate state before reaching equilibrium, offering a quantitative atomic-scale framework for understanding and designing phase transformations in complex multielement nanomaterial systems.

**Acknowledgments:** The authors thank the Analysis Center at Tsinghua University for the use of aberration-corrected electron microscopes. This work is also supported by High-performance Computing Platform of Peking University.

**Funding:** This work was supported by Beijing Natural Science Foundation (Grant No. F251007), the National Natural Science Foundation of China (Grant No. 92477203) and the National Key R&D Program of China (Grant No. 2024YFA1509500).

**Author contributions:** J.Z. and J.X. conceived the idea, and J.Z directed the study. J.X. conducted the four-dimensional atomic-resolution ET data collection experiments under the supervision of J.Z. J.X. and D.J. synthesized the Pd@Ir core-shell NPs. J.X. and Z. S. completed in situ STEM tests. J.X. and Y.D. completed the gas set up required for the experiment, J.X. implemented the annealing processes to the specimen under specific atmosphere. J.X. and Y.Z. discussed the methods of atomic-resolution ET reconstruction. J.X. and Z.L. discussed the calculation in curvature. J.X. analyzed the 3D atomic structures and interpreted the results under the supervision of J.Z. J.X. and J.Z. wrote the manuscript. All authors commented on the manuscript.

**Competing interests:** Authors declare that they have no competing interests.

**Data and materials availability:** All data are available in the main text or the supplementary materials.

**Supplementary Materials**

Materials and Methods

Figs. S1 to S32

Tables S1 and S3

References (*43-54*)

Movie S1

# Supplementary Materials for

## Tracking atomic-scale interdiffusion in immiscible bimetallic nanoparticles via four-dimensional electron tomography

Jisheng Xie, Dijin Jiang, Zhen Sun, Yiheng Dai, Zezhou Li, Yao Zhang, Jihan Zhou*

*Corresponding author: Jihan Zhou, jhzhou@pku.edu.cn

**The PDF file includes:**

Materials and Methods
Figs. S1 to S32
Tables S1 and S3

**Other Supplementary Material for this manuscript include the following:**

Movies S1

## Materials and Methods

1. Chemicals

Sodium tetrachloropalladate(II) ($Na_2PdCl_4$, ≥99.99%) was purchased from Sigma-Aldrich; iridium (III) acetylacetonate ($Ir(acac)_3$) was purchased from Bide Pharmatech, Shanghai, China; ethanol, cyclohexane and acetone were purchased from TGchem, Beijing, China; potassium bromide (KBr, 99.7%) was purchased from MREDA, Beijing, China; potassium chloride (KCl, 99.7%) and L-ascorbic acid (99.7%) were purchased from Xilong Scientific, Guangdong, China; Poly(vinyl-pyrrolidone) (PVP, MW ≈ 55000 g/mol) was purchased from Yuanye, Shanghai, China; benzylalcohol (96%, Cat No.1014772) and oleylamine (90%, Cat No. 1017269) were purchased from Leyan, Shanghai, China. All the chemicals were used as received without further purification.

2. Sample preparation

Pd@Ir NPs were synthesized by a modified two-step method described in detail elsewhere (*34, 43*).

1) *Synthesis of Pd NPs*

PVP (315 mg), KBr (15 mg), KCl (610.5 mg) and L-ascorbic acid (180 mg) were dissolved in 24 mL of deionized water, and the solution was heated to 80°C. $Na_2PdCl_4$ (30 mg) was dissolved in 9 mL deionized water, sonicated for 10 min, and then combined with the preheated solution. The mixture was stirred at 80°C for 4 h, and allowed to cool to room temperature. After the reaction, 5 mL of ethanol and 33 mL of acetone were added, and the product was collected by centrifugation at 10000 rpm for 2 min. The resulting Pd seeds were redispersed in 6 mL of ethanol-water mixture (1:1, v:v), and sonicated for 5 min for further use.

2) *Synthesis of Pd@Ir NPs*

The Pd seeds dispersion obtained above (4.7 mL) was first mixed with 6 mL of benzyl alcohol and stirred at 200°C in an open vessel for 5 min to complete phase transfer of the seeds from ethanol to benzyl alcohol. The resulting solution was then cooled to room temperature. An Ir precursor solution was prepared by dissolving 30.6 mg of $Ir(acac)_3$ and 62.5 mg of PVP in 6.5 mL of benzyl alcohol with sonication until fully dissolved. The precursor solution was combined with the phase-transferred Pd seed solution, transferred into a Teflon-lined autoclave, and heated at 200°C for 12 h; it was then cooled naturally to room temperature. The as-obtained mixture was further mixed with 5 mL of oleylamine, reacted again at 200°C for 50 min, and cooled to room temperature. An ethanol/acetone (1:1, v:v) mixed solution was added, and the product was isolated by centrifugation at 10000 rpm for 2 min. The final Pd@Ir nanoparticles were redispersed in ethanol and sonicated for 5 min before further use.

3. Material characterizations

Annular dark-field scanning transmission electron microscopy (ADF-STEM), energy-dispersive X-ray spectroscopy (EDS) and in situ STEM during heating experiments were carried out on an aberration-corrected FEI Themis Z microscope at 300 kV. In situ heating was performed using a Chip-Nova volcano series in situ ultra-heating single-tilt holder. ADF-STEM images were acquired after a holding time of approximately 5 min once the target temperature was reached, to compensate for thermal drift.

3. Four-dimensional atomic-resolution tomography

1) *Specimen preparation*

A single-slot silicon TEM window with a $SiN_x$ membrane (purchased from NORCADA, Canada) was used for the specimen preparation. To further enhance the conductivity of the membrane, a 4-nm-thick carbon layer was deposited onto the TEM window using a sputter coater (pulse mode, Leica EM ACE600). The grid was then heated at 350 ℃ for 12 h in an argon-filled glove box to remove contamination. Pd@Ir nanoparticles dispersed in ethanol were sprayed onto the $SiN_x$ membrane using an atomizer. The grid was subsequently dried in the vacuum, heated at 250 ℃ in argon for 4 h to remove the majority of the surface surfactant (*34*), and stored in an argon-filled glove box for the further use.

2) *4D Tomography experiment*

Three Pd@Ir NPs were randomly selected for the 4D tomography experiment. To obtain NPs in different diffusion states, the three NPs were annealed under an argon flow at 200, 300, and 400°C, respectively. For each particle, four tomographic tilt series were collected at cumulative annealing times of 0, 30, 60, and 120 min (figs. S4-15). The detailed annealing procedure was as follows (fig. S19):

(a) The first tomographic tilt series of Particle-1 was acquired on a prepared specimen prepared as described in step (1).

(b) The specimen was placed on a quartz boat (60×30×15 cm), transferred into a tubular furnace, and annealed at 200°C for 30 min.

(c) After annealing, the second tomographic tilt series of the same Particle-1 was acquired.

(d): Step (b) and (c) was repeated to get the third tomographic tilt series.

(e) The specimen was then annealed twice as in the step (b) and the fourth tomographic tilt series for Particle-1 was acquired. (f) Steps (a)-(e) were repeated at 300°C and 400°C to get 4D tomographic tilt series for Particles-2 and 3, respectively.

3) *Operation parameters used in acquisitions of tomographic tilt series*

All tomographic tilt series were acquired in ADF-STEM mode on an aberration-corrected FEI Themis Z microscope. The operation parameters were: accelerating voltage 300 kV, convergence semi-angle 25 mrad, HAADF detector inner semi-angle 40.6 mrad and HAADF detector outer semi-angle 200 mrad, and a pixel size of 0.352 Å. To minimize sample drift, three images' frames at each tilt angle were acquired with a dwell time of 2 μs. A relatively low electron dose rate was used for each tomographic tilt series, measured to be 2.6-3.0 × $10^5$ $e^-$ $Å^{-2}$ (Supplementary Tables 1-3). This dose is less than 40% of that reported in ref. (*31*), which effectively protected the NPs from beam damage. We compared all experimental zero-degree images with forward projections calculated from the final 3D reconstructions (*44*), and all comparisons showed no noticeable structural changes or beam damage in Particles 1-3 (figs. S16-18).

## 4. 3D atomic-resolution reconstruction

1) *Image pre-processing*

For each tilt series, the images were pre-processed prior to reconstruction following an established procedure (*27, 39*). First, at each tilt angle, the three acquired frames were registered by normalized cross-correlation and then averaged to improve the signal-to-noise ratio; during this step, linear drifts were estimated and corrected. Subsequently, noise was removed from the registered images using a block-matching and 3D filtering algorithm (*45*). After denoising, a mask of the NP was generated for each image using Otsu thresholding (*46*). To preserve surface details in the tomographic reconstruction, a buffer zone of approximately 10 pixels was retained between the NP and the mask edge. The background intensity inside the mask was estimated by Laplace interpolation and then subtracted. Finally, all images in each tilt series were aligned to sub-pixel accuracy using center-of-mass and common-line alignment methods.

2) *3D atomic-resolution reconstruction*

The Real Space Iterative Reconstruction (RESIRE) algorithm (*44*) was employed for the reconstruction, with the specific parameters listed in Supplementary Tables 1-3. To correct the residual misalignment and angular inaccuracies caused by stage and holder instability, spatial realignment and angular refinement were performed as suggested in other AET experiments (*23, 31, 36*). Using the refined projections and angles, the final 3D volume was generated using the RESIRE algorithm.

5. Determination of 3D atomic coordinates and chemical species

Using the final 3D volume obtained above, a series of steps were performed to determine the atomic coordinates and identify the chemical species.

1) *Identification of local intensity maxima*

A polynomial fitting method (*26, 47*) was applied within a voxel region of 7×7×7 box to locate peak locations. From the preliminary list, local maxima in the reconstructed density were selected as candidate atomic positions, with a minimum separation of 2.2 Å enforced between adjacent candidates. Three-dimensional polynomial fitting was then performed on each candidate to achieve sub-pixel precision. This process automatically identified and finalized over 98% of the atomic positions. Owing to the "missing wedge" problem (*48, 49*), the reconstructions contain artifacts and noise, which occasionally lead to inaccurate fitting or false atomic sites. For a small subset of atoms (fewer than 2%), adjustments were made in cases where local intensity maxima were physically to close or an expected atom was missing within an intensity blob (*23, 39*).

2) *Element classification*

Initial element classification was conducted using the K-means clustering algorithm (*50*). For each atomic coordinate, the integrated intensity was calculated by summing the intensity signal within a 7×7×7 box centered on the atom. These integrated intensities were then grouped into two classes: Pd (lower intensity) and Ir (higher intensity). Because of the "missing wedge" problem (*48, 49*), surface atoms exhibit systematically lower intensities compared to interior atoms, particularly along the missing wedge direction. To mitigate this intensity variation, we employed a combination of local re-classification (*28*) and atom flipping (*31*) methods, as described below:

(a) The integrated intensities of surface atoms were re-classified separately by K-means algorithm to suppress the "missing wedge" effect.

(b) For the same NP at different annealing times, we identified corresponding atom pairs between neighboring datasets (detailed in the next section). Most atom pairs shared the same chemical species, and only a small fraction exhibited different species.

(c) From this minority, we randomly selected an atom and flipped its species (Pd to Ir, or Ir to Pd). Based on the atomic models before and after atom flipping, we computed two sets of multislice-simulated projections across all tilt angles (*51*). Because flipping a single atom affects only a local region in the projection space, the simulation was restricted to that local region to accelerate computation.

(d) The errors between the simulated and experimental projections were quantified, and the species assignment was updated if it reduced the error.

(e) Steps (b)-(d) were repeated for all atom pairs with initially mismatched species until the assignments stabilized.

6. Structural analysis of final atomic models

1) *Calculation of CN, curvature, specific surface area, and atomic distance to surface*

Pair distribution functions (PDF) were computed from the final atomic models. CNs were determined by setting the first minima of the PDFs as the cutoff distance. The atoms with CN<12 were classified as surface atoms. The local curvature at each surface atom was calculated following a previously reported method (*52*). The specific surface area was quantified by generating a surface mesh using the alpha-shape algorithm (*53*), with the probe sphere radius taken as the position of the first peak in the PDFs. Similarly, The distances of each atom to the surface were calculated as its shortest distance to the surface mesh.

2) *Identification of common positions and chemically consistent atoms*

First, an ideal FCC lattice was fitted to each dataset by a least-squares fitting approach (*20*), matching all atomic positions well. Next, we compared the two atomic models obtained from the same NP at different annealing times. Atomic pairs were identified by requiring that the two atoms (one atom from each dataset) deviated by less than half the distance of the first minimum in PDFs (*31*). These matched pairs are referred to as common positions (Supplementary Tables 1-3). Atoms that did not form pairs were classified as either disappeared positions or new positions in the corresponding dataset. Subsequently, we evaluated the chemical consistency of the paired atoms. Most of the common-position pairs retained the same chemical species and are identified as chemically consistent atoms. Migrated-inconsistent atoms were defined as those occupying common positions but having changed their chemical identity. Overall, the set of migrated atoms comprised atoms from disappeared positions, atoms form new positions and migrated-inconsistent-atoms.

3) *Quantification of chemical short-range-order parameter (CSROP)*

To evaluate short-range order in each NP, a pairwise multicomponent short-range-order parameter was applied (*54*), which is denoted as:

$$\alpha_{ij}^{m} = \left(p_{ij}^{m} - C_j\right) / \left(\delta_{ij} - C_j\right)$$

Here, *m* indicates the *m*-th nearest-neighbor shell around a central atom $i$, $p_{ij}^{m}$ represents the averaged probability of detecting a *j*-type atom in the *m*-th shell around a central atom $i$, $C_j$ is the system-wide concentration of *j*-type atoms, and $\delta_{ij}$ is the Kronecker delta function. In this work, we used the first shell (m=1) to calculate $\alpha_{Pd-Pd}$ and $\alpha_{Ir-Pd}$. For homoatomic pairs, such as $\alpha_{Pd-Pd}$, a positive value implies a tendency toward Pd segregation within the shell, whereas a negative value indicates mixing. For heteroatomic pairs, such as $\alpha_{Ir-Pd}$, a negative value suggests clustering of Pd atoms around a central Ir atom, and a positive value reflects dispersion. The parameters $\alpha_{Pd-Pd}$ and -$\alpha_{Ir-Pd}$ values were therefore taken as indicators of Pd-segregation: positive values for both metrics point to a preference for Pd-rich local environments. To follow the evolution of Pd segregation at different annealing times, surface atoms were grouped, based on the cuboctahedral morphology of NPs, into eight regions with {111} orientation and six regions with {100} orientation. And the CSROPs were averaged over each region. The resulting region-averaged CSROPs were tracked across different annealing times, and the differences were computed and designated as ΔCSROPs. To monitor the inward diffusion of Ir, each NP was peeled layer-by-layer from the surface inward (like peeling an onion), the atomic compositions of each layer was determined, and CSROP was averaged within each layer to quantify the evolution of Ir abundance (Fig. 3C).

4) *Calculation of interfacial diffusion coefficients of Ir atoms*

The radial concentration profiles of Ir at different depths were quantified based on the “onion peeling” method described in the previous section (fig. S32). To eliminate interference from particle reshaping during surface reconstruction, we selected the changes in Ir concentrations with the core-shell interfaces (the region within 6 Å of the surface) before and after the 2nd and 3rd annealing steps for further analysis. The diffusion coefficients were calculated according to Fick’s second law (*41*):

$$\frac{\partial C_{Ir}}{\partial t} = D \frac{\partial^2 C_{Ir}}{\partial x^2}$$

where D is the interfacial diffusion coefficient, $C_{Ir}$ is the Ir concentration in the respective layer, $t$ is the annealing time, and $x$ is the interlayer spacing.

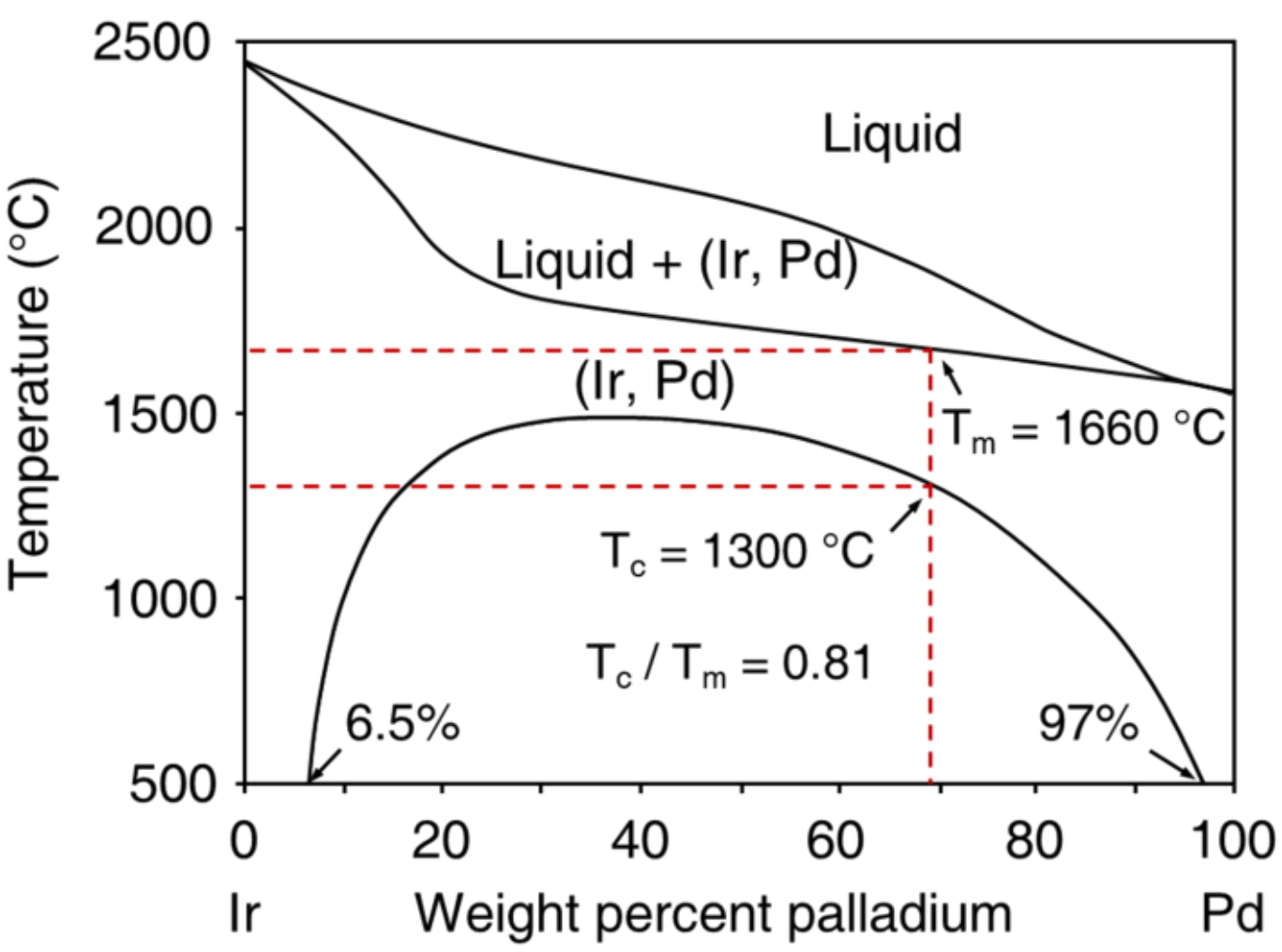

**Fig. S1. Bulk phase diagram of the Pd–Ir binary system (*33*).**

The red dashed line indicates $T_c$ and $T_m$ in the bulk material, respectively, which are corresponded to the Pd-Ir stoichiometric ratio of Pd@Ir NPs (69 wt% for Pd) used in this study. $T_c/T_m$ = 1573/1933K = 0.81.

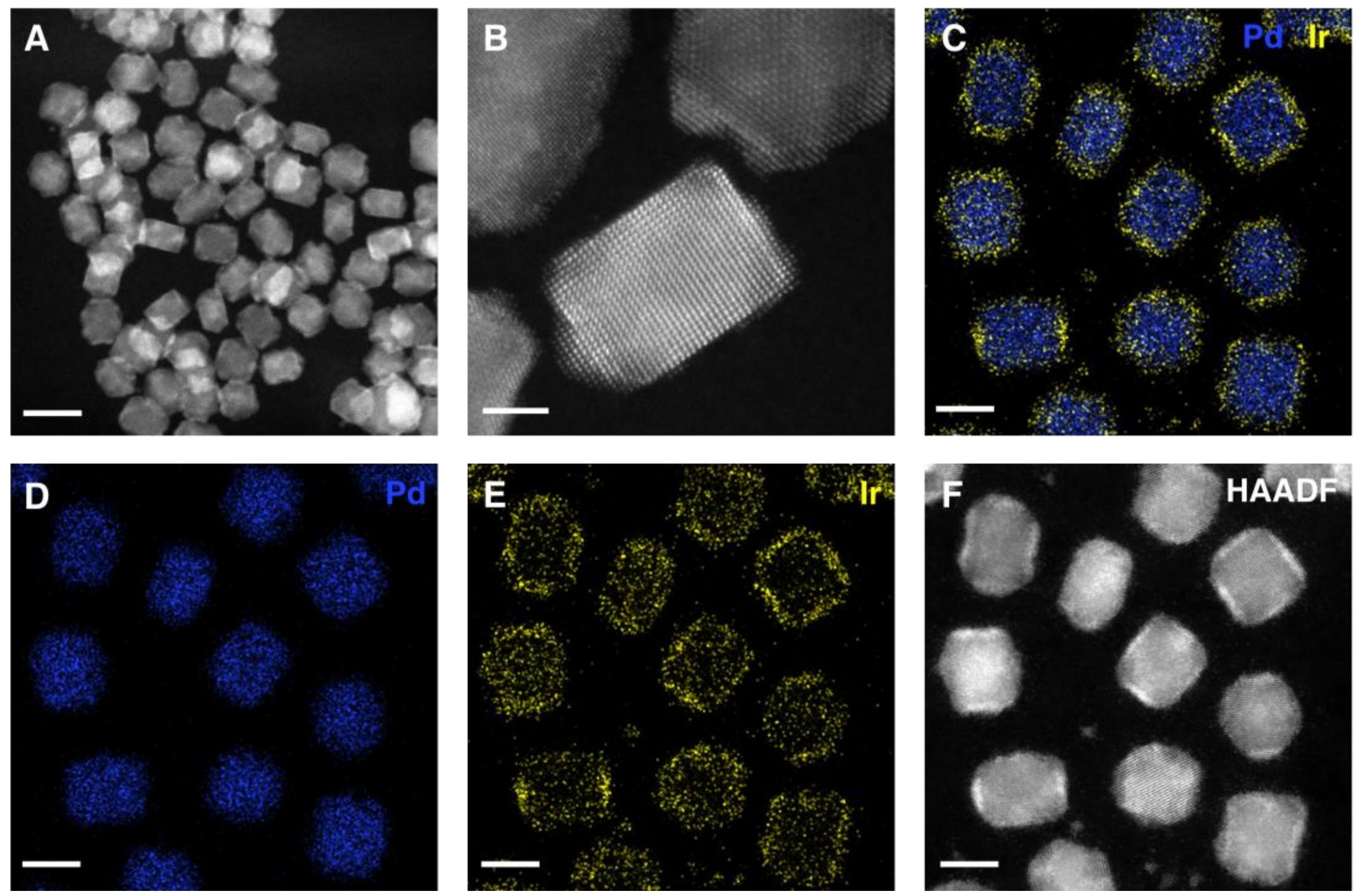

**Fig. S2. Characterization of Pd@Ir NPs.**

(**A** and **B**) ADF-STEM images of Pd@Ir NPs at low (A) and high (B) magnifications. (**C** to **F**) EDS characterization for Pd@Ir NPs. Scale bars are 10 nm in (A), 2 nm in (B), and 5 nm in (C to F) respectively.

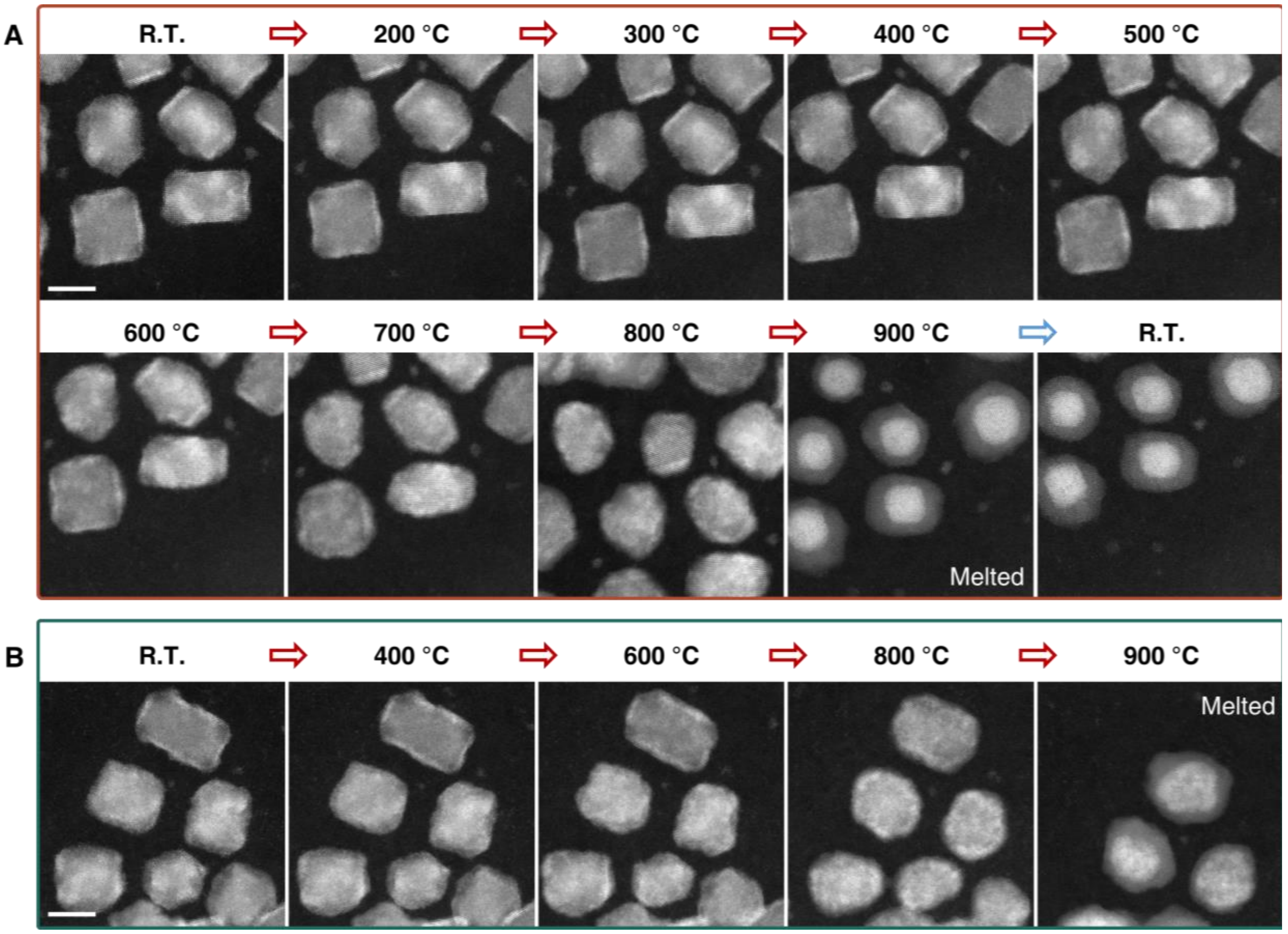

**Fig. S3. In situ heating experiments for Pd@Ir NPs.**

(**A** and **B**) ADF-STEM images of Pd@Ir NPs at different temperatures for the same field of views. Scale bars are 5 nm for all images. To investigate the effects of different temperatures, ADF-STEM imaging was performed after a 5-min holding time when the temperature reaches the corresponding number.

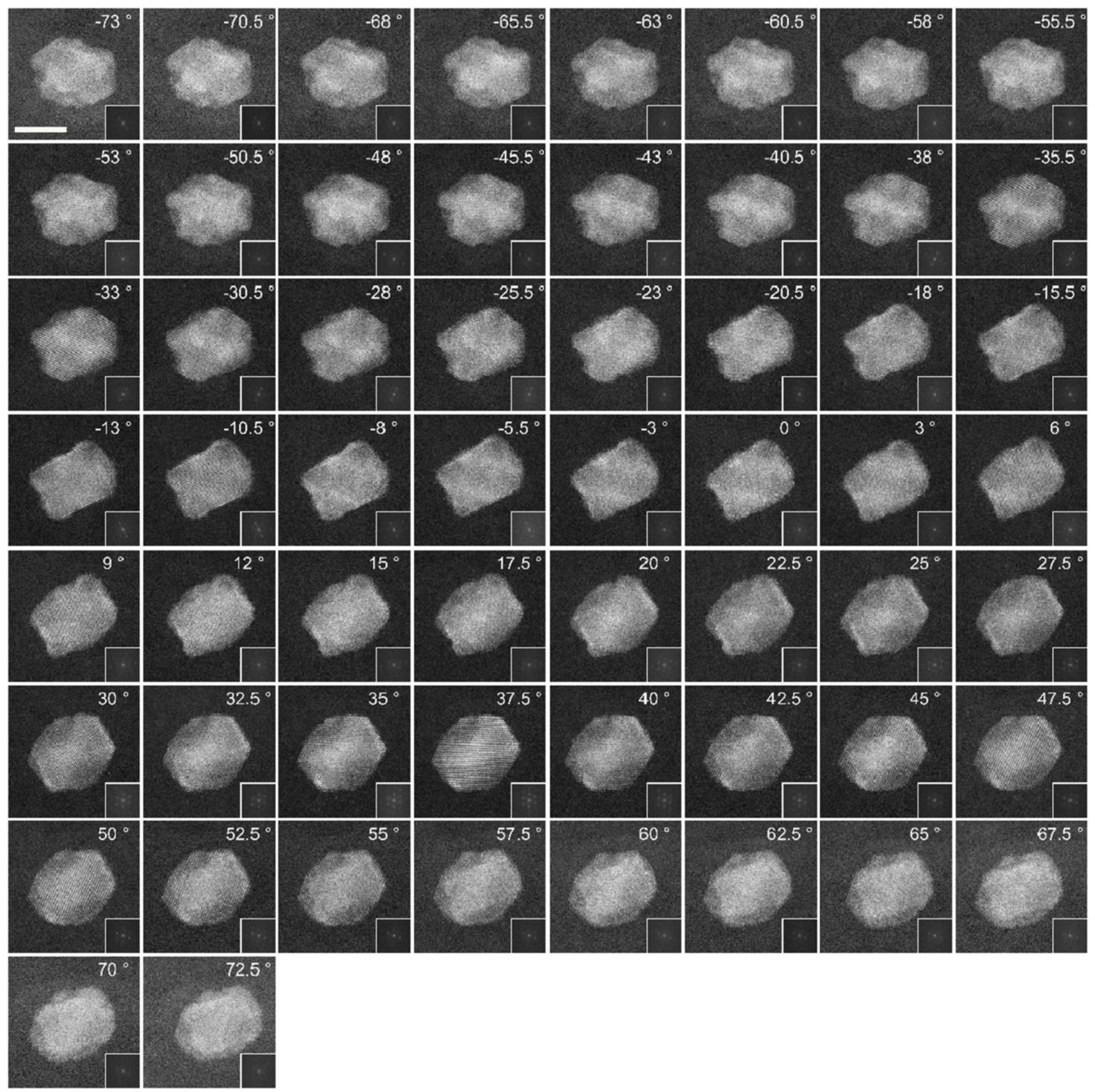

**Fig. S4. Tomographic tilt series of Pd@$Ir_{200_0}$ NP.**

ADF-STEM images with a tilting range from −73.0° to +72.5°. The value of angle to which the image belongs is recorded in the upper right corner of each image. Corresponding patterns of fast fourier transform (FFT) are shown in the insets. Scale bar is 5 nm.

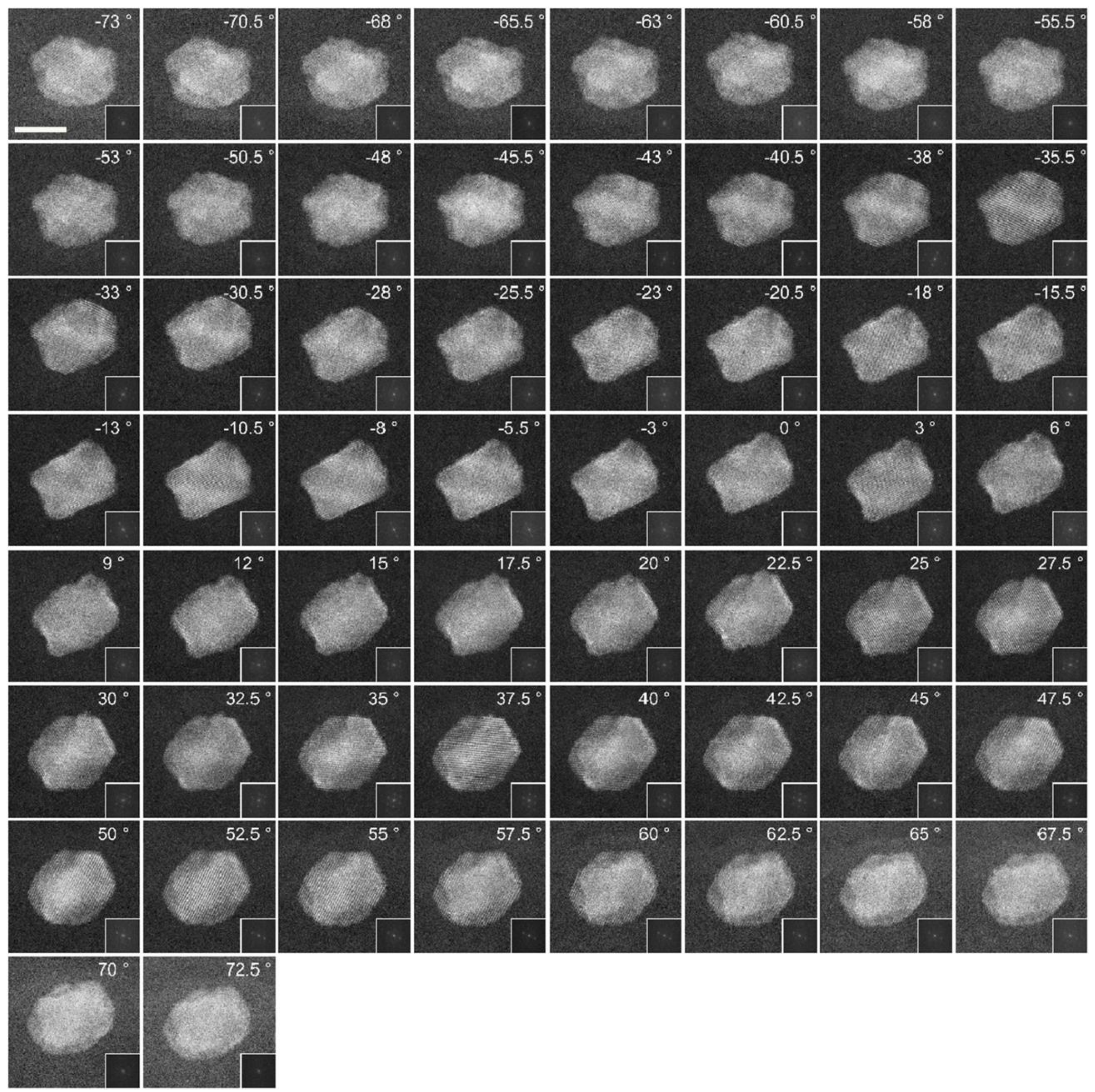

**Fig. S5. Tomographic tilt series of $Pd@Ir_{200_1}$ NP.**

ADF-STEM images with a tilting range from −73.0° to +72.5°. The value of angle to which the image belongs is recorded in the upper right corner of each image. Corresponding FFT patterns are shown in the insets. Scale bar is 5 nm.

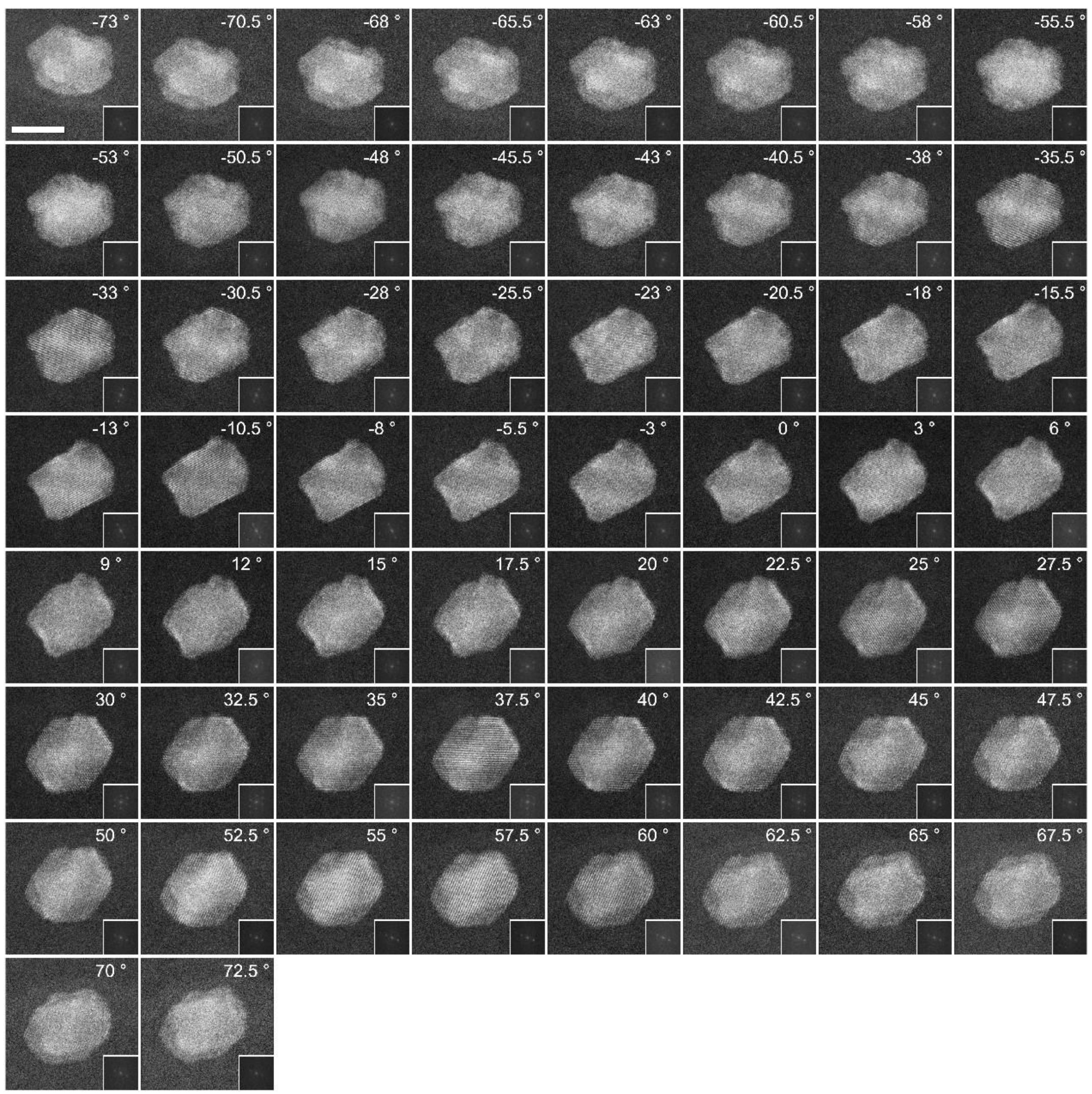

## Fig. S6. Tomographic tilt series of $Pd@Ir_{200_2}$ NP.

ADF-STEM images with a tilting range from −73.0° to +72.5°. The value of angle to which the image belongs is recorded in the upper right corner of each image. Corresponding FFT patterns are shown in the insets. Scale bar is 5 nm.

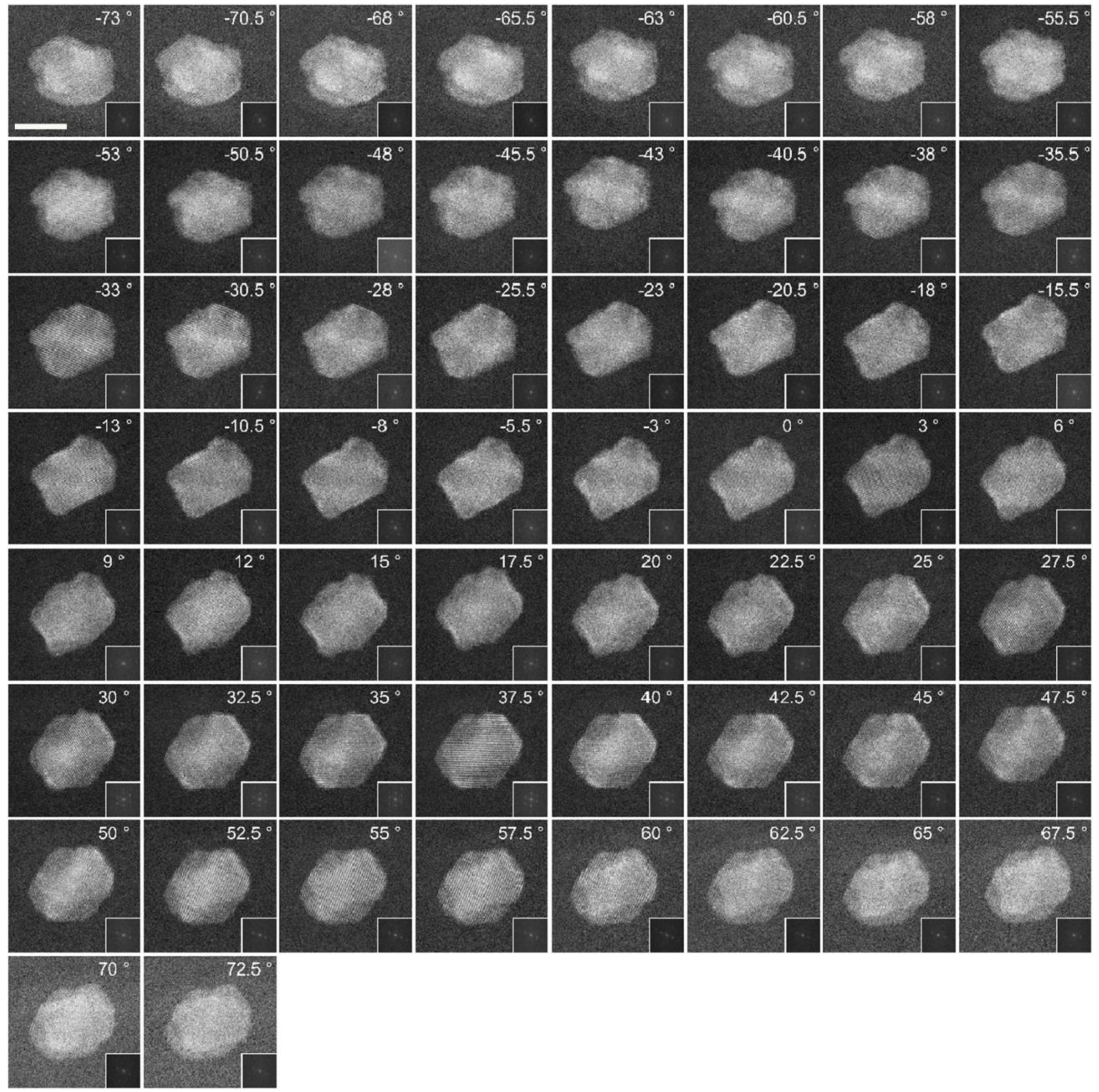

**Fig. S7. Tomographic tilt series of $Pd@Ir_{200_3}$ NP.**

ADF-STEM images with a tilting range from −73.0° to +72.5°. The value of angle to which the image belongs is recorded in the upper right corner of each image. Corresponding FFT patterns are shown in the insets. Scale bar is 5 nm.

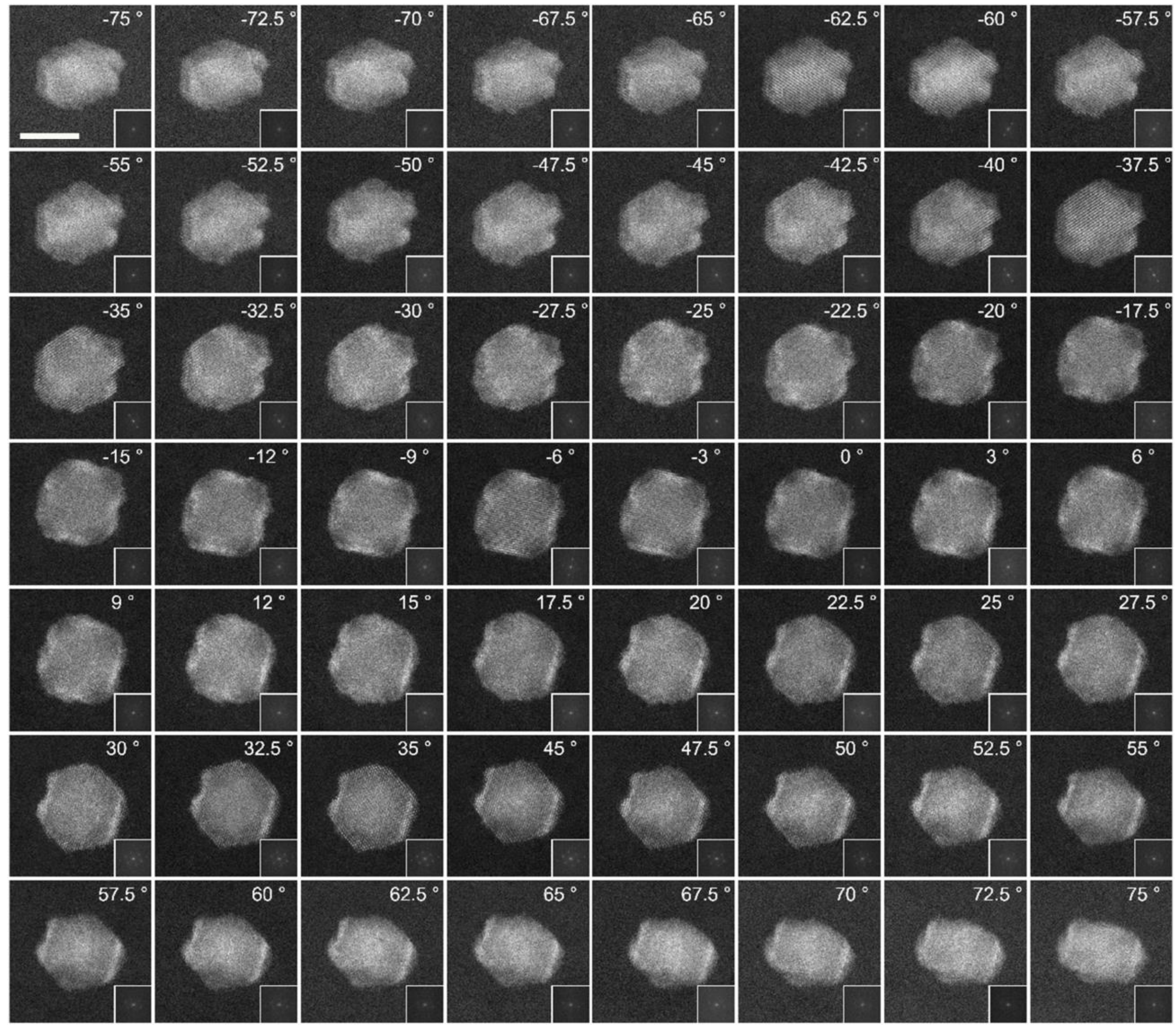

**Fig. S8. Tomographic tilt series of Pd@$Ir_{300_0}$ NP.**

ADF-STEM images with a tilting range from −75.0° to +75.0°. The value of angle to which the image belongs is recorded in the upper right corner of each image. Corresponding FFT patterns are shown in the insets. Scale bar is 5 nm.

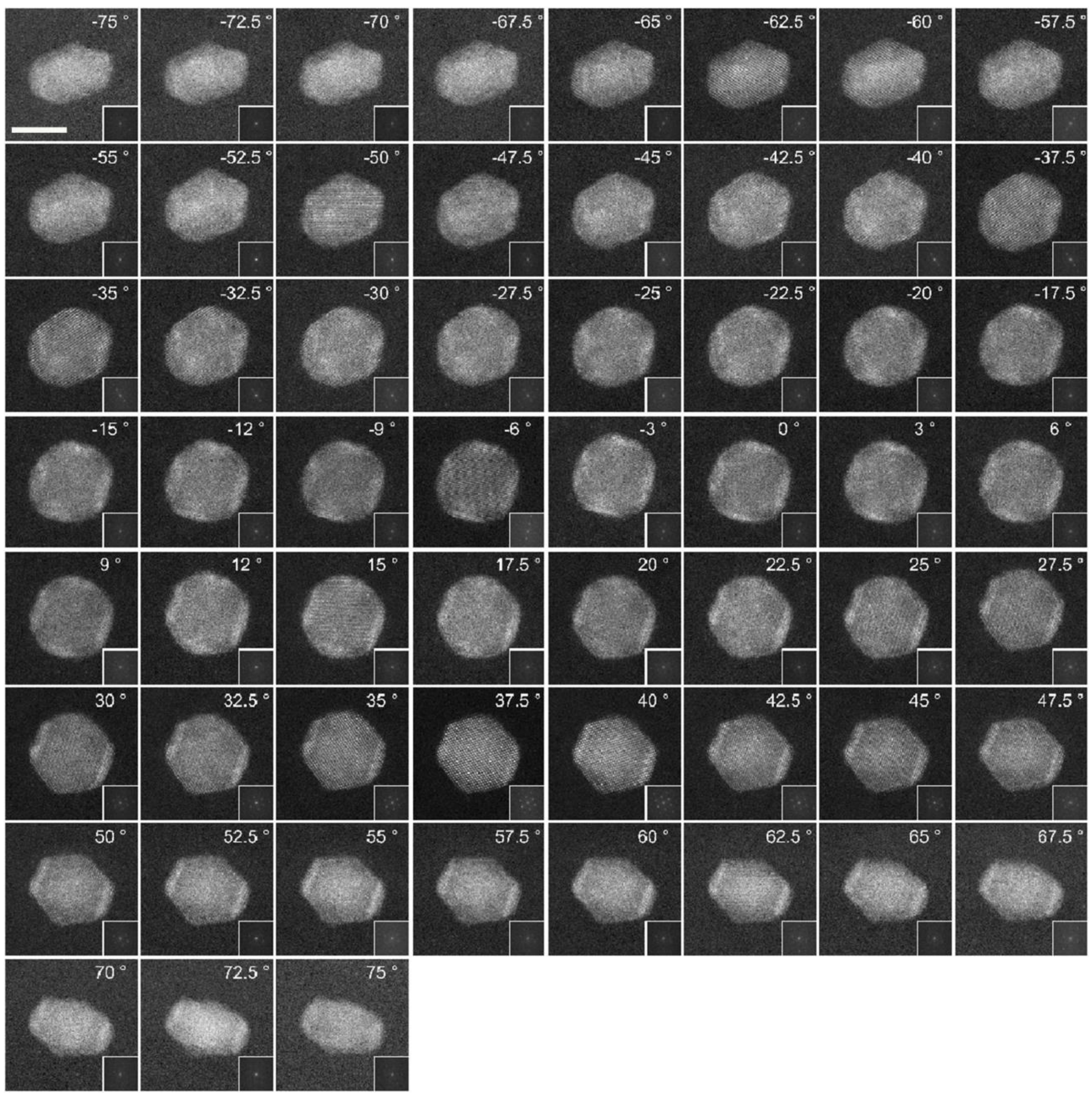

**Fig. S9. Tomographic tilt series of Pd@Ir$_{300_1}$ NP.**

ADF-STEM images with a tilting range from −75.0° to +75.0°. The value of angle to which the image belongs is recorded in the upper right corner of each image. Corresponding FFT patterns are shown in the insets. Scale bar is 5 nm.

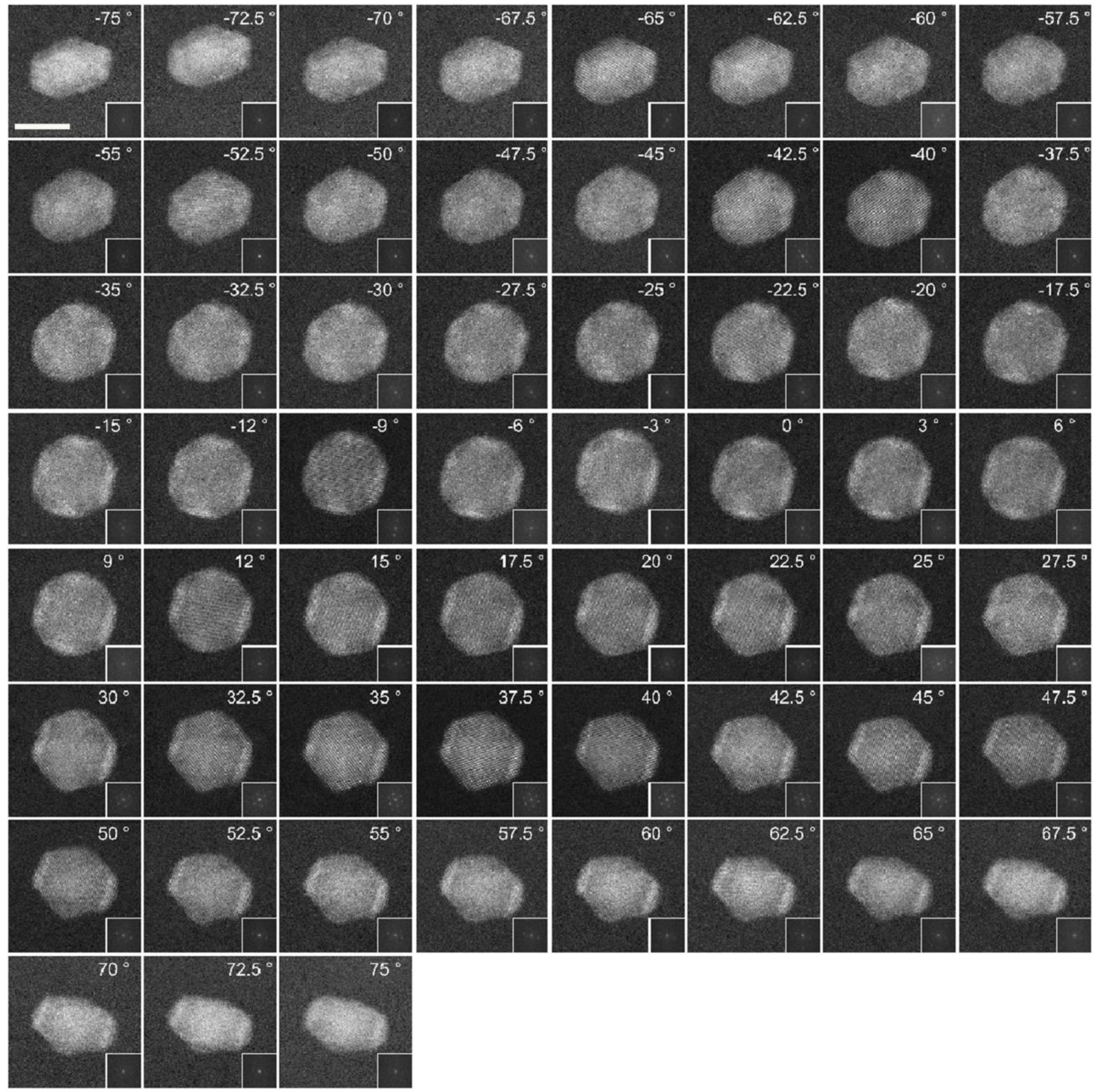

## Fig. S10. Tomographic tilt series of Pd@$Ir_{300_2}$ NP.

ADF-STEM images with a tilting range from −75.0° to +75.0°. The value of angle to which the image belongs is recorded in the upper right corner of each image. Corresponding FFT patterns are shown in the insets. Scale bar is 5 nm.

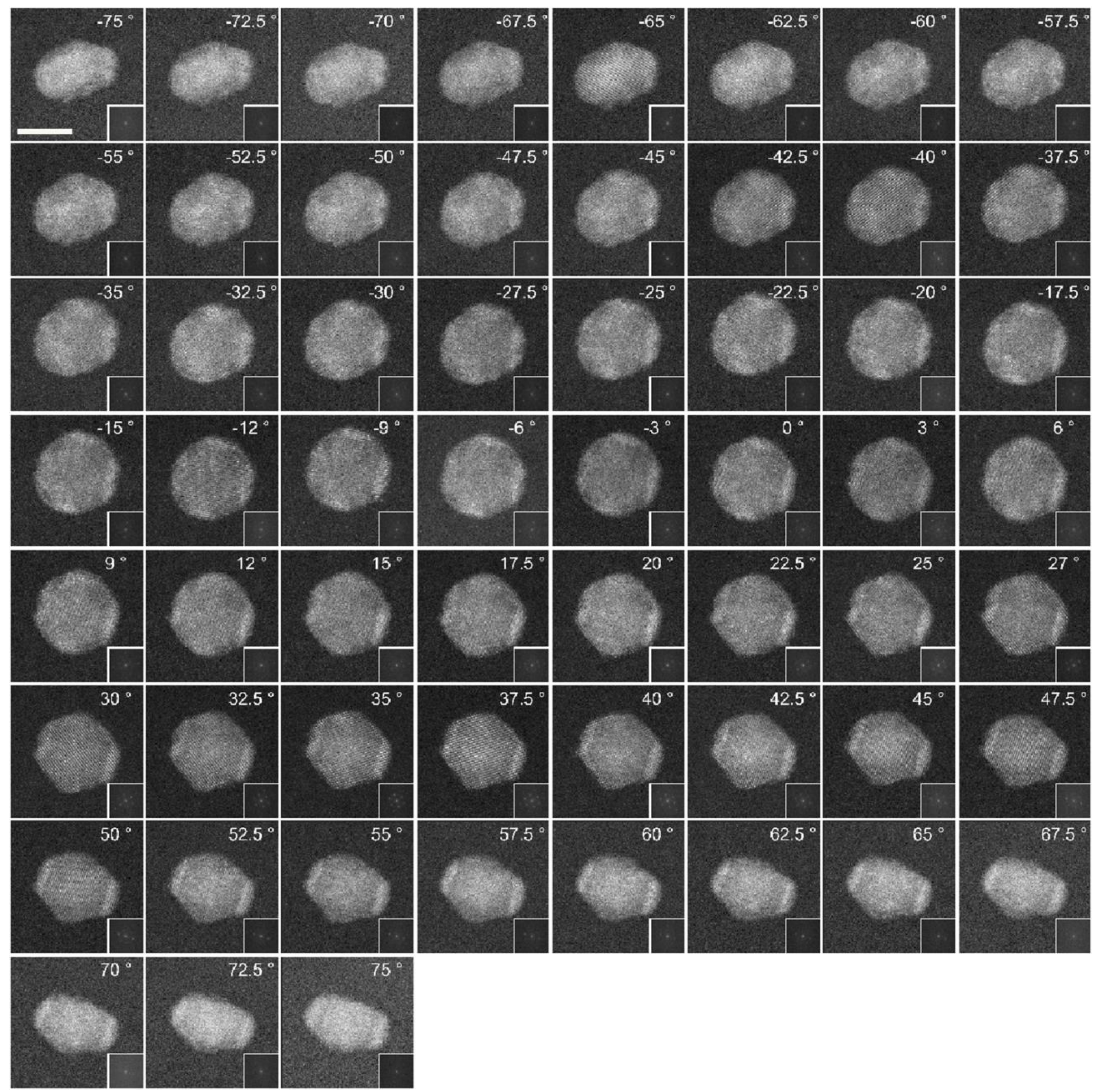

**Fig. S11. Tomographic tilt series of $Pd@Ir_{300_3}$ NP.**

ADF-STEM images with a tilting range from −75.0° to +75.0°. The value of angle to which the image belongs is recorded in the upper right corner of each image. Corresponding FFT patterns are shown in the insets. Scale bar is 5 nm.

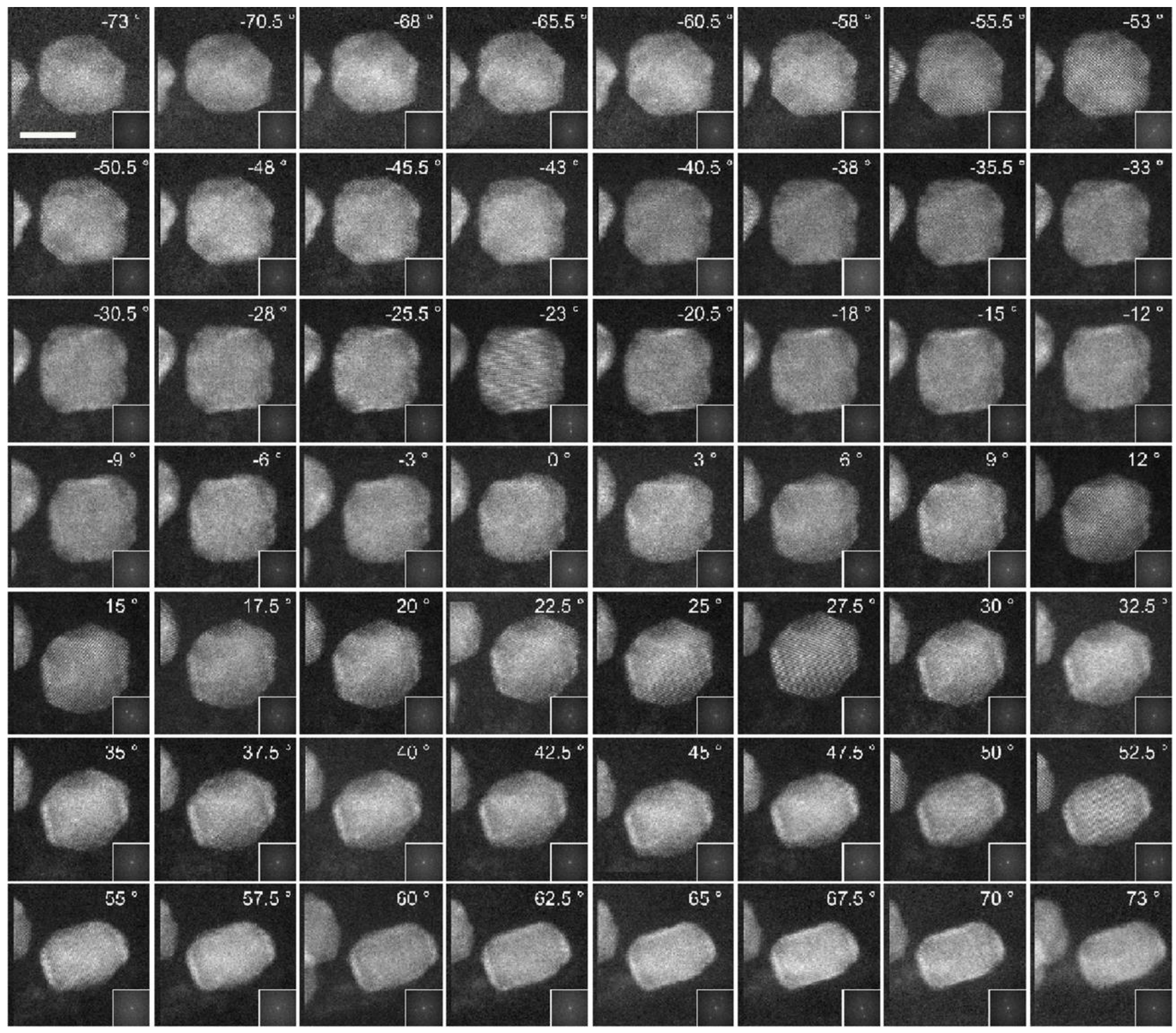

**Fig. S12. Tomographic tilt series of Pd@Ir$_{400_0}$ NP.**

ADF-STEM images with a tilting range from −73.0° to +73.0°. The value of angle to which the image belongs is recorded in the upper right corner of each image. Corresponding FFT patterns are shown in the insets. Scale bar is 5 nm.

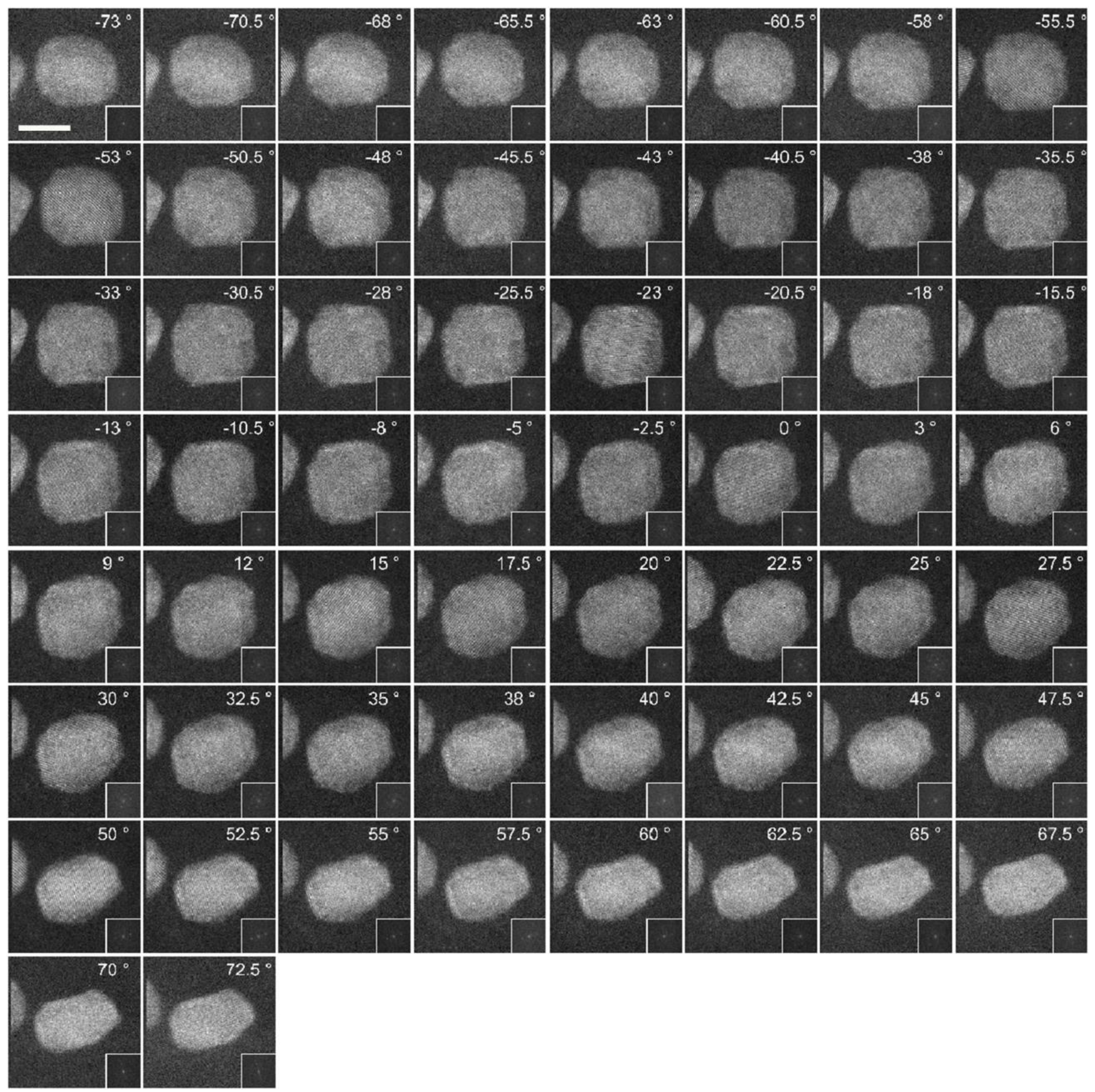

**Fig. S13. Tomographic tilt series of Pd@Ir$_{400_1}$ NP.**

ADF-STEM images with a tilting range from −73.0° to +72.5°. The value of angle to which the image belongs is recorded in the upper right corner of each image. Corresponding FFT patterns are shown in the insets. Scale bar is 5 nm.

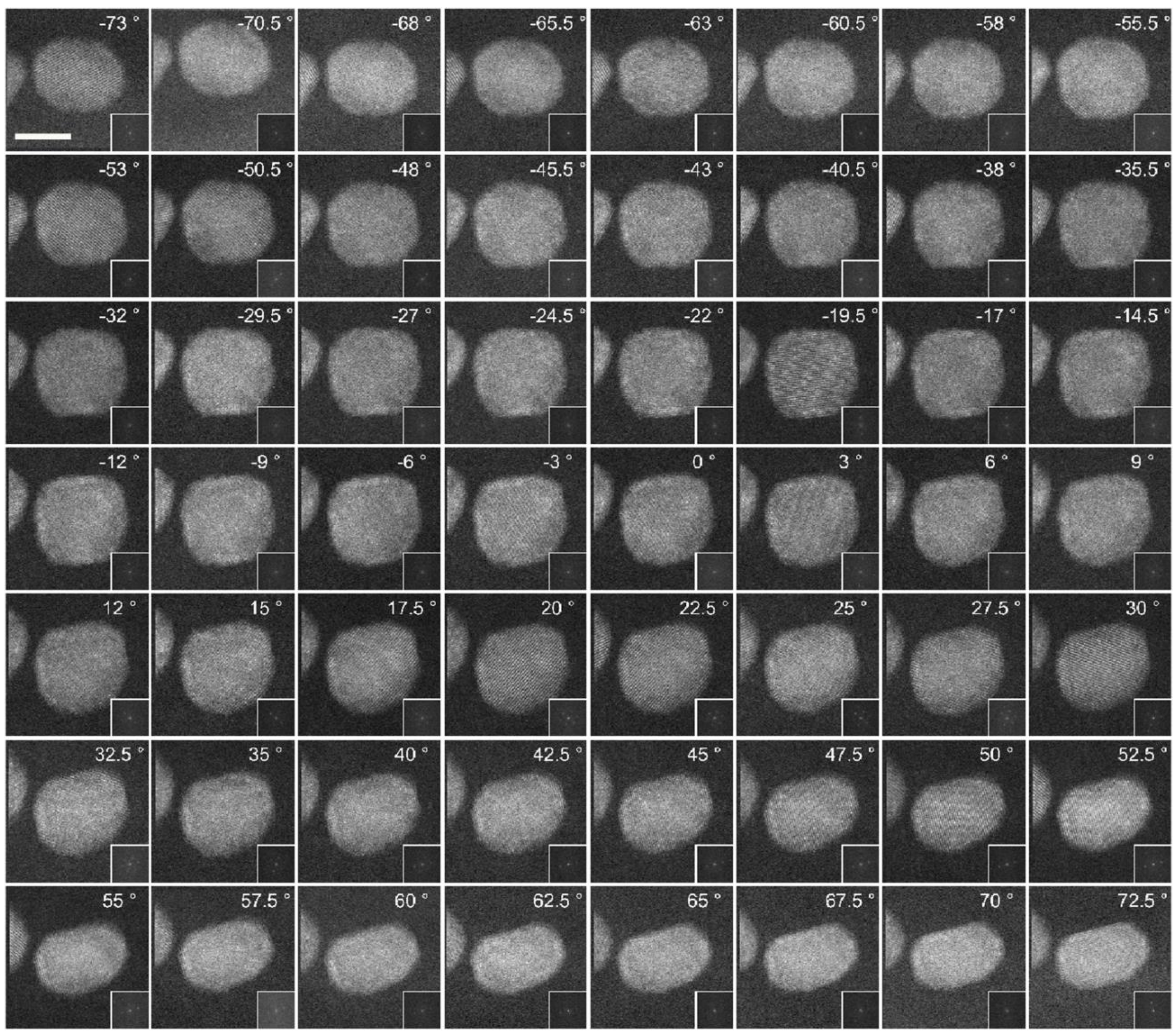

**Fig. S14. Tomographic tilt series of $Pd@Ir_{400_2}$ NP.**

ADF-STEM images with a tilting range from −73.0° to +72.5°. The value of angle to which the image belongs is recorded in the upper right corner of each image. Corresponding FFT patterns are shown in the insets. Scale bar is 5 nm.

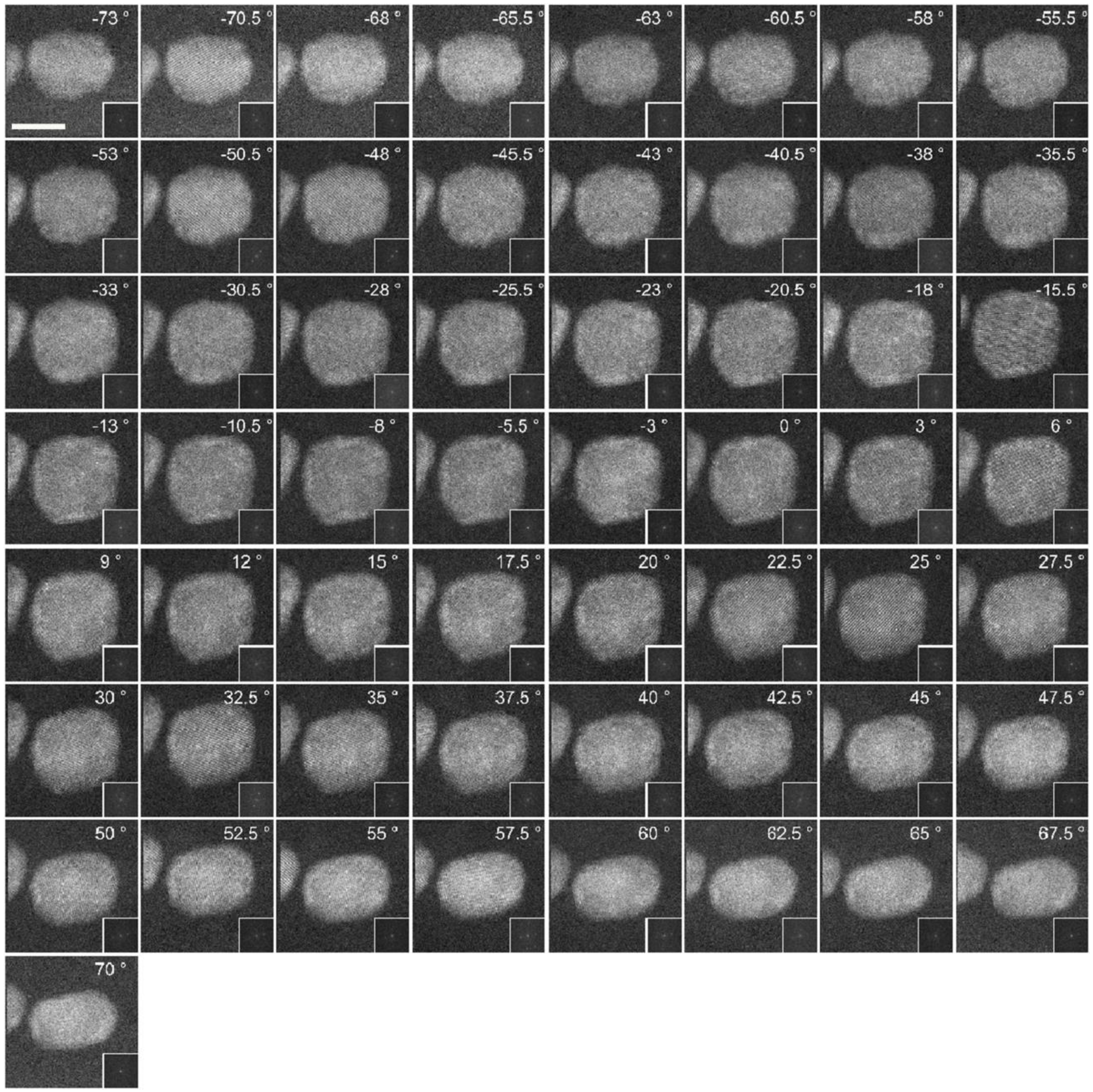

**Fig. S15. Tomographic tilt series of Pd@Ir$_{400_3}$ NP.**

ADF-STEM images with a tilting range from −73.0° to +70.0°. The value of angle to which the image belongs is recorded in the upper right corner of each image. Corresponding FFT patterns are shown in the insets. Scale bar is 5 nm.

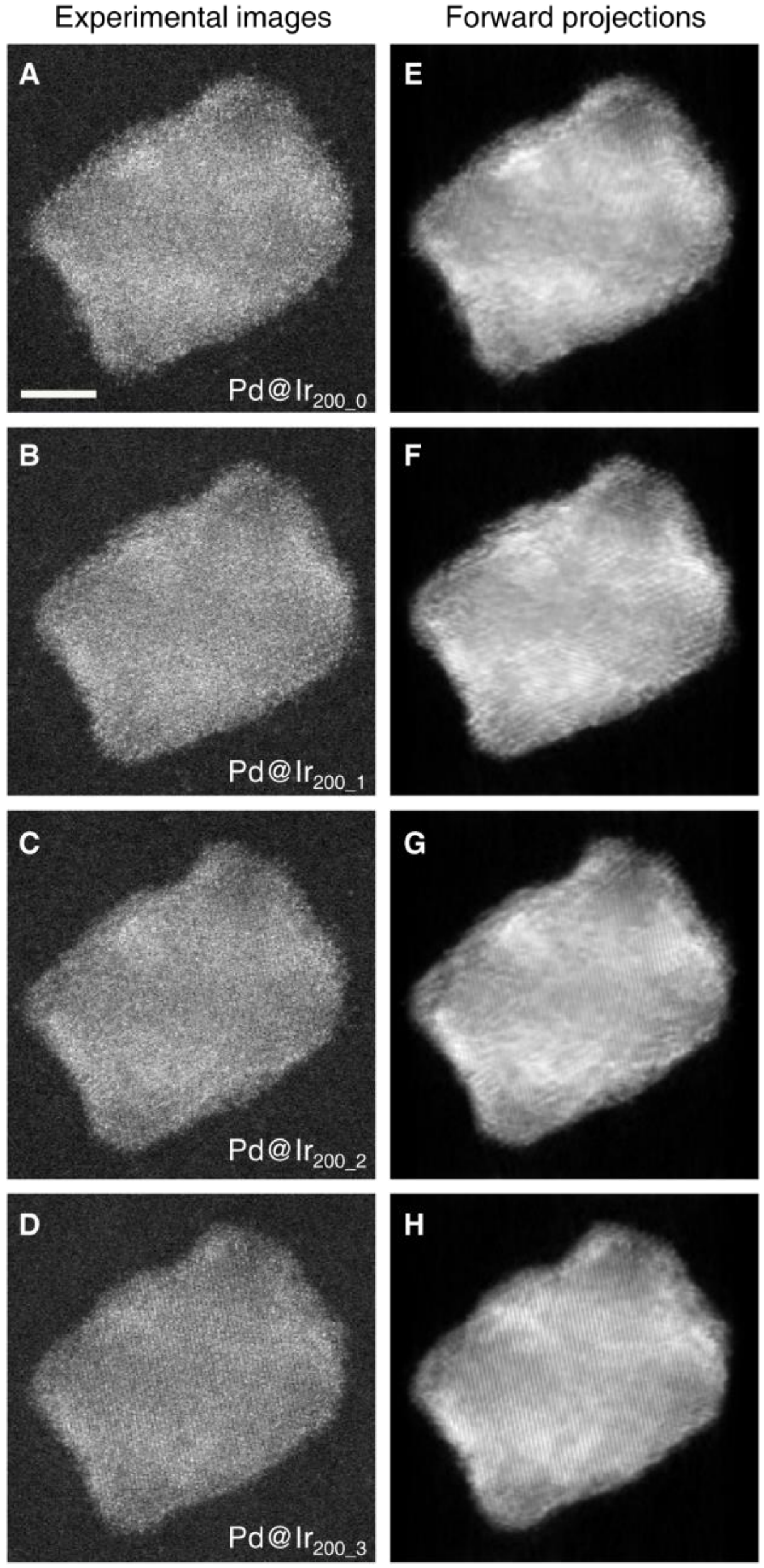

**Fig. S16. Consistency check of Particle-1 series.**

(**A** to **D**) ADF-STEM images taken at 0° during tilting experiment for (A) Pd@$Ir_{200_0}$, (B) Pd@$Ir_{200_1}$, (C) Pd@$Ir_{200_2}$ and (D) Pd@$Ir_{200_3}$, respectively. (**E** to **H**) Simulated forward projections of the final 3D atomic model at the tilt angles for (E) Pd@$Ir_{200_0}$, (F) Pd@$Ir_{200_1}$, (G) Pd@$Ir_{200_2}$ and (H) Pd@$Ir_{200_3}$, respectively. Scale bars for (A) to (H) are 2 nm.

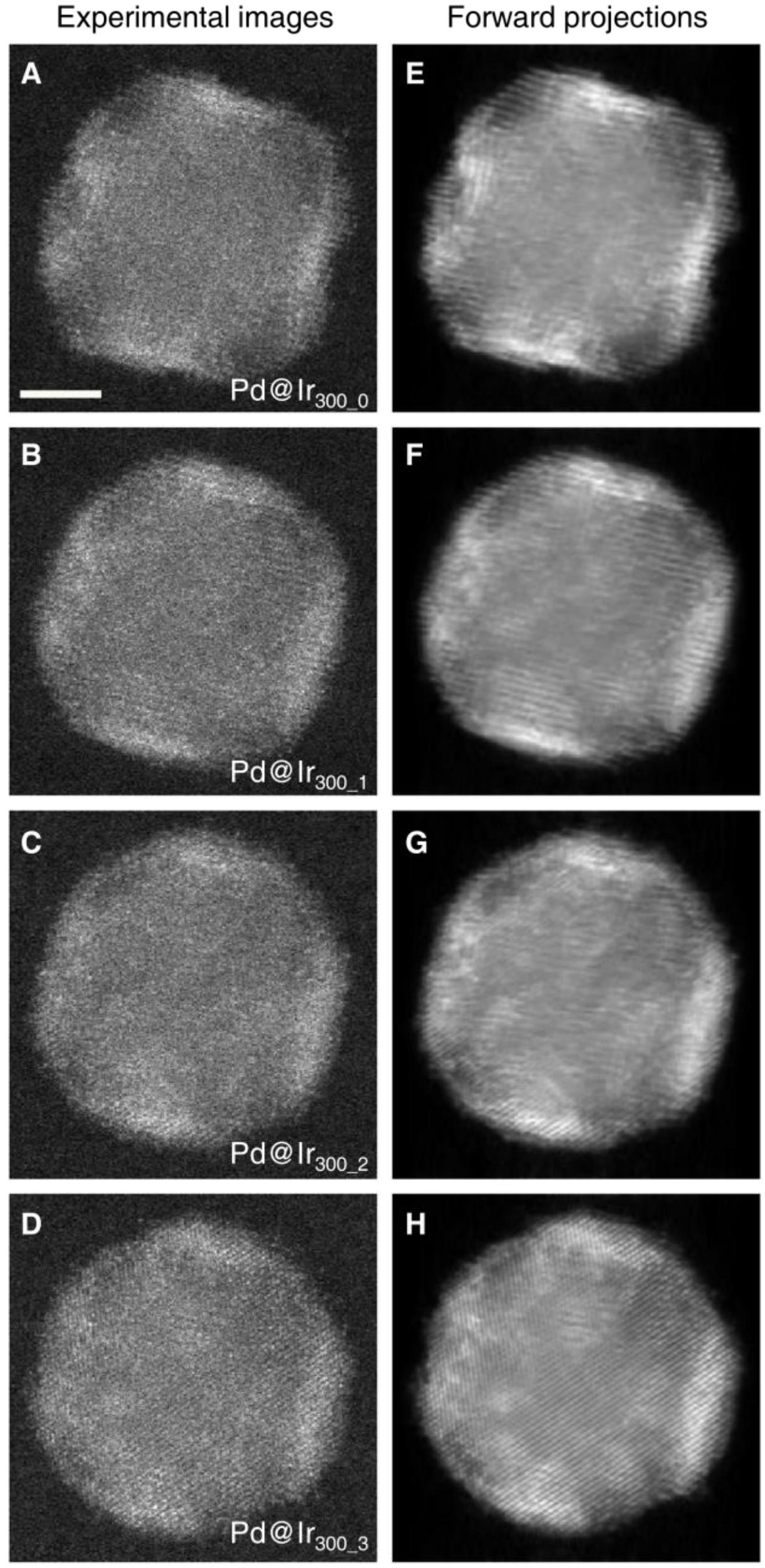

**Fig. S17. Consistency check of Particle-2 series.**

(**A** to **D**) ADF-STEM images taken at 0° during tilting experiment for (A) Pd@$Ir_{300_0}$, (B) Pd@$Ir_{300_1}$, (C) Pd@$Ir_{300_2}$ and (D) Pd@$Ir_{300_3}$, respectively. (**E** to **H**) Simulated forward projections of the final 3D atomic model at the tilt angles for (E) Pd@$Ir_{300_0}$, (F) Pd@$Ir_{300_1}$, (G) Pd@$Ir_{300_2}$ and (H) Pd@$Ir_{300_3}$, respectively. Scale bars for (A) to (H) are 2 nm.

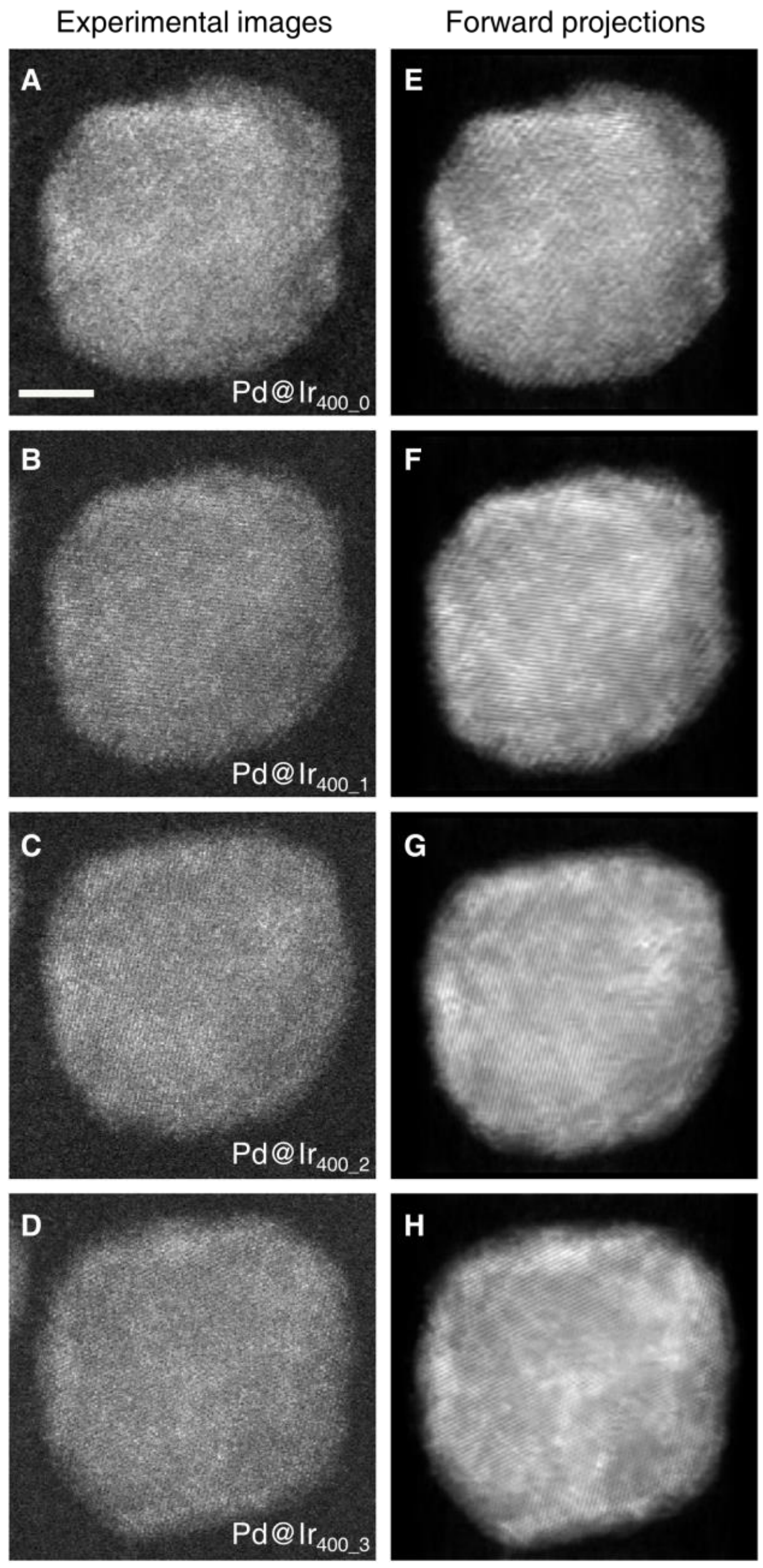

**Fig. S18. Consistency check of Particle-3 series.**

(**A** to **D**) ADF-STEM images taken at 0° during tilting experiment for (A) $Pd@Ir_{400_0}$, (B) $Pd@Ir_{400_1}$, (C) $Pd@Ir_{400_2}$ and (D) $Pd@Ir_{400_3}$, respectively. (**E** to **H**) Simulated forward projections of the final 3D atomic model at the tilt angles for (E) $Pd@Ir_{400_0}$, (F) $Pd@Ir_{400_1}$, (G) $Pd@Ir_{400_2}$ and (H) $Pd@Ir_{400_3}$, respectively. Scale bars for (A) to (H) are 2 nm.

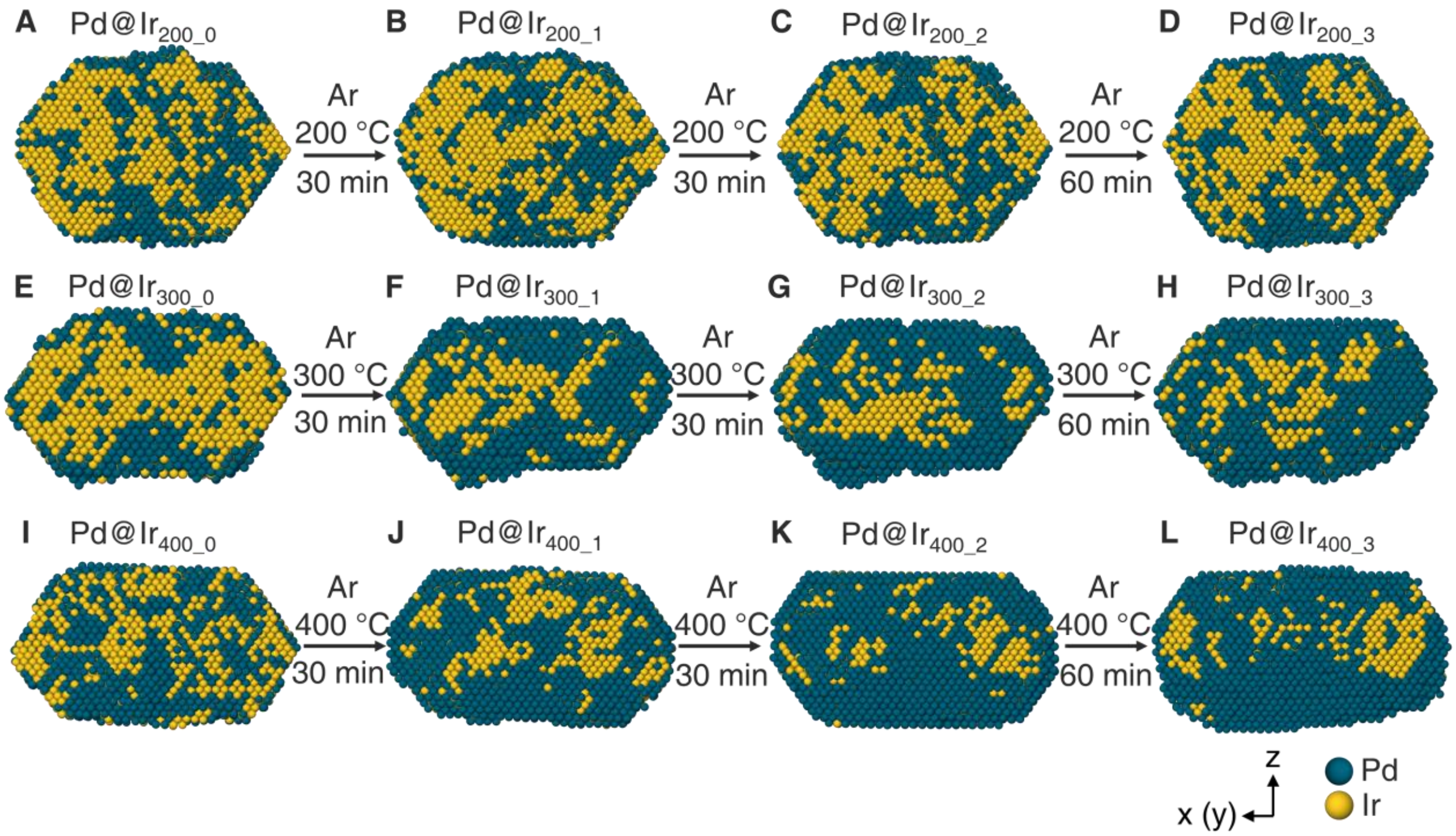

**Fig. S19. The 3D atomic models of Particle 1-3.**

(**A** to **D**) 3D atomic models for Particle-1 after 0-min (A), 30-min (B), 60-min (C) and 120-min (D) annealing, respectively. (**E** to **H**) 3D atomic models for Particle-2 after 0-min (E), 30-min (F), 60-min (G) and 120-min (H) annealing, respectively. (**I** to **L**) 3D atomic models for Particle-3 after 0-min (I), 30-min (J), 60-min (K) and 120-min (L) annealing, respectively.

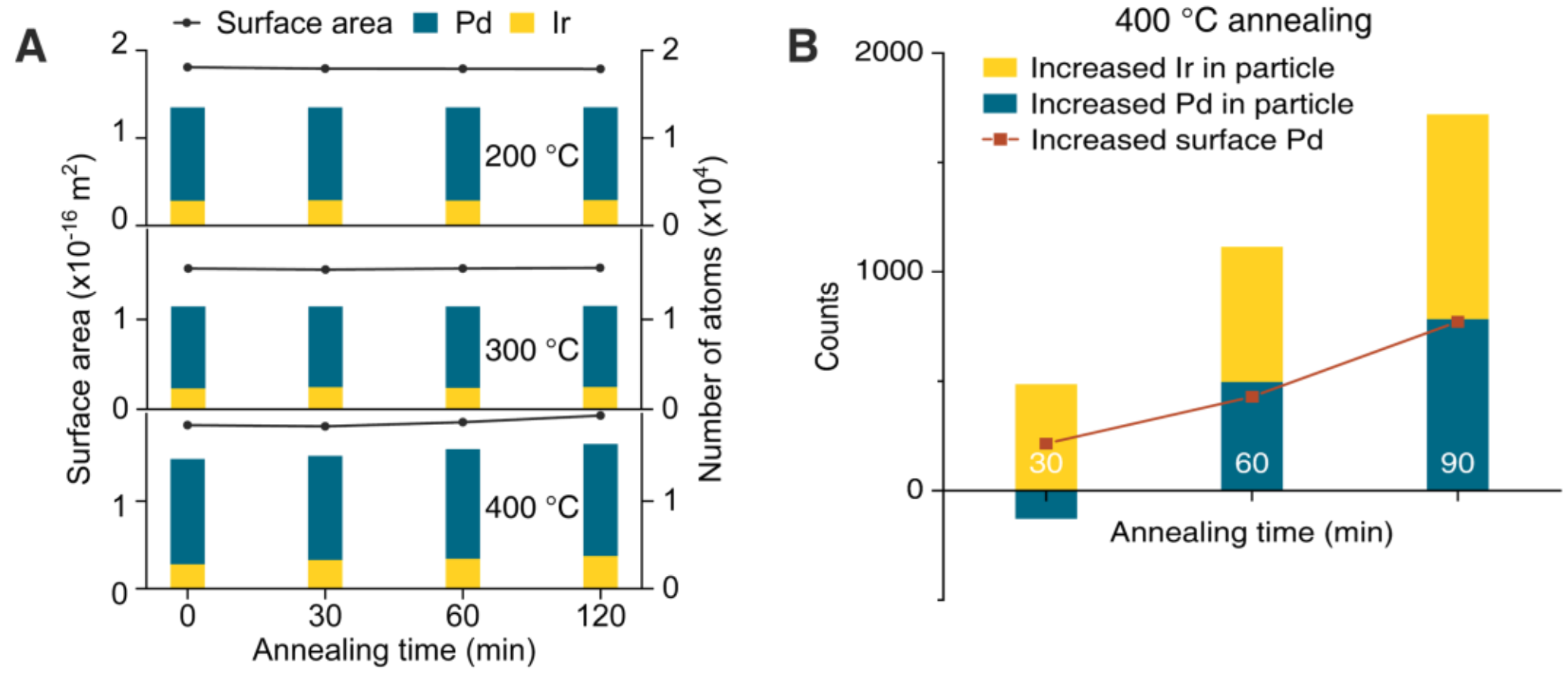

**Fig. S20. General structure analysis for Particle 1-3.**

(**A**) Atom counts and surface area. (**B**) Atom counts in particle and surface ranges for Particle-3 (annealing at 400°C).

While the increased atoms in particle range (due to Oswald ripening) are mainly composed of Ir atoms, surface Pd atoms show an increasing trend during the annealing process **(b)**. This indicates that the Pd-rich surface (fig. S19) is largely due to Pd segregation, instead of Oswald ripening at this temperature.

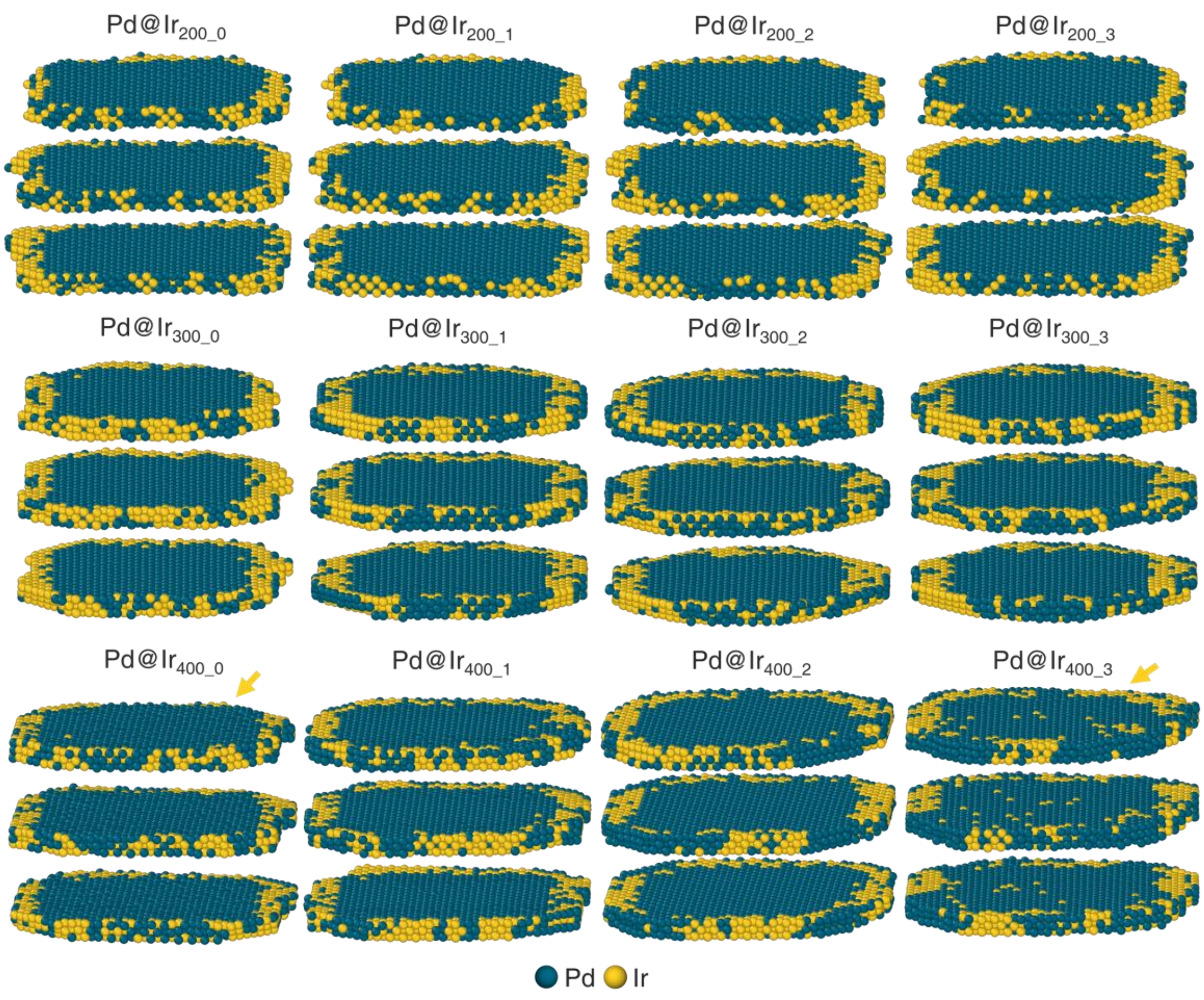

**Fig. S21. Atomic slices of Particles 1-3.**

Green and yellow represent Pd and Ir, respectively. Yellow arrows point to the Ir inward diffusion at 400°C.

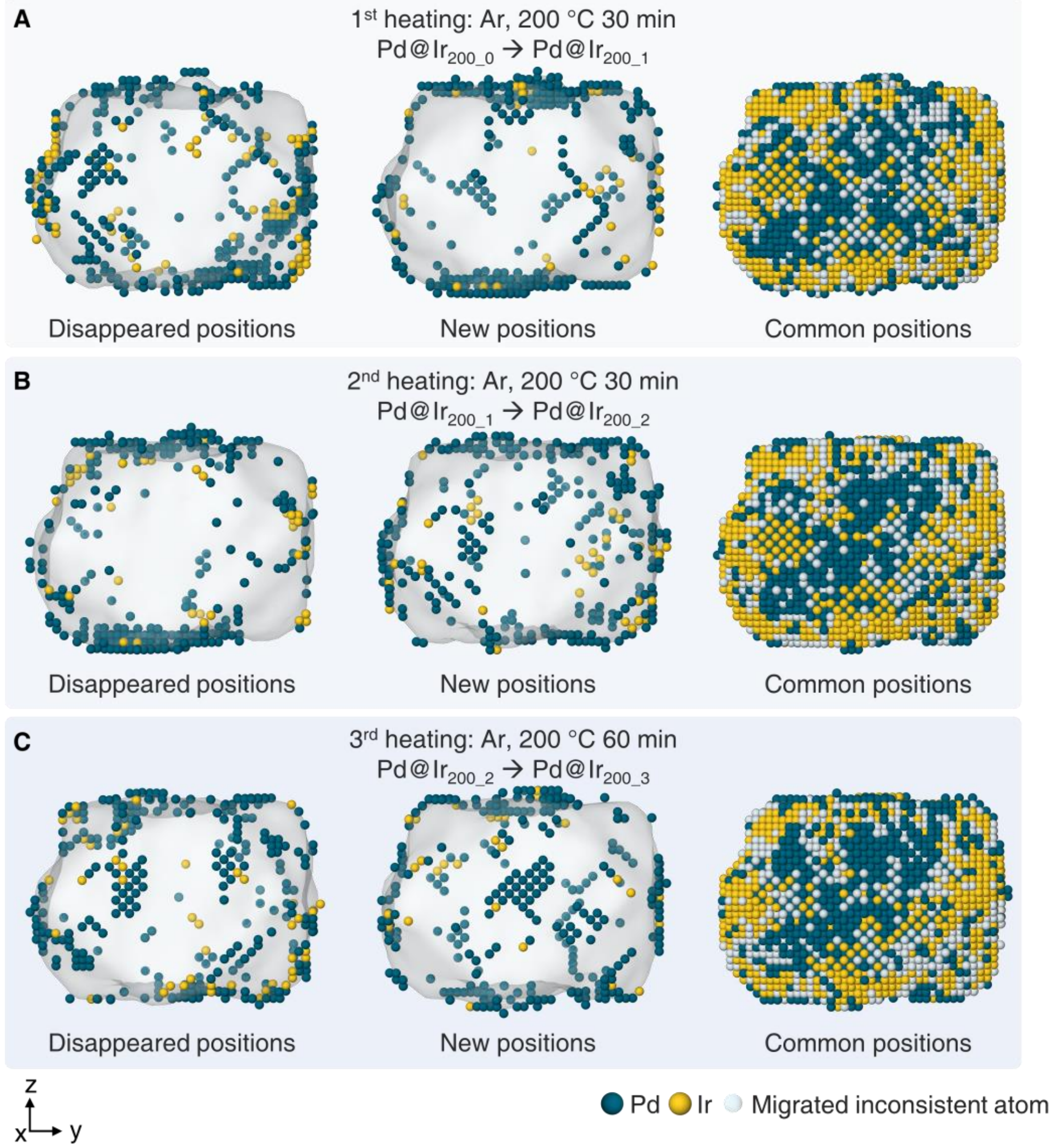

**Fig. S22. 4D evolution of atomic positions and chemical species in Particle-1.**

(**A**) 3D renderings of Particle-1 before and after the 1st annealing (30-min annealing in Ar). (**B**) 3D renderings of Particle-1 before and after the 2nd annealing (60-min annealing in Ar). (**C**) 3D renderings of Particle-1 before and after the 3rd annealing (120-min annealing in Ar for two times). Green, yellow, and white spheres represent Pd, Ir, and migrated-inconsistent-atoms at the common positions, respectively.

The left panels display disappeared positions, the positions identified in the earlier stage but absent in the later stage. The central panels show new positions, the positions that appear only in the later stage. The right panels present common positions, which are conserved positions across both stages. Because potential atom exchanges can occur at common positions (reflecting volume diffusion within the particle), we define migrated-inconsistent atoms as those occupying common positions but exhibiting a change in chemical identity. Migrated atoms are composed of disappeared positions, new positions and migrated-inconsistent-atoms.

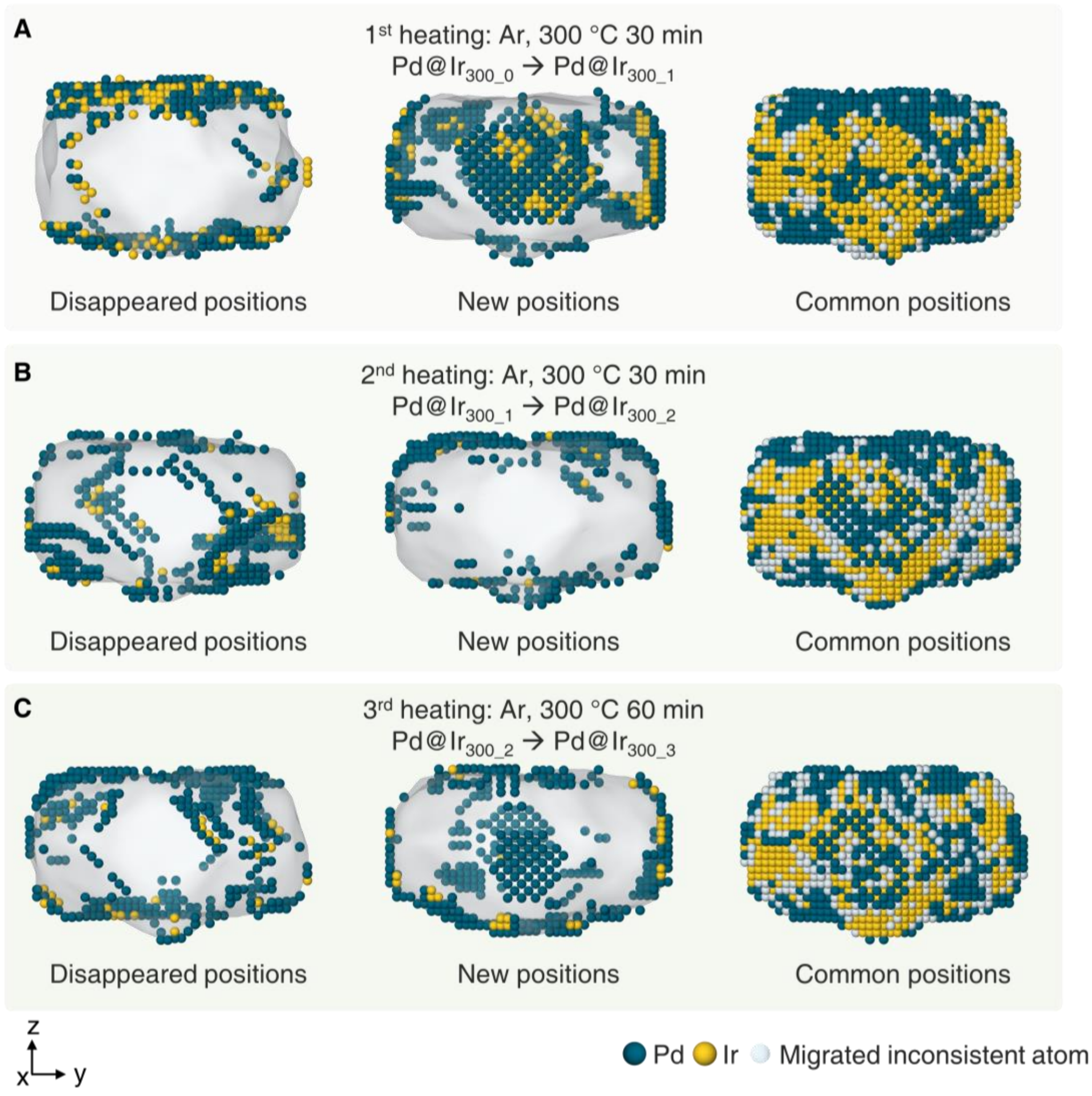

**Fig. S23. 4D evolution of atomic positions and chemical species in Particle-2.**

(**A**) 3D renderings of Particle-2 before and after the 1st annealing (30-min annealing in Ar). (**B**) 3D renderings of Particle-2 before and after the 2nd annealing (60-min annealing in Ar). (**C**) 3D renderings of Particle-2 before and after the 3rd annealing (120-min annealing in Ar for two times).

The definitions of disappeared positions, new positions, common positions, migrated-inconsistent atoms, and migrated atoms are provided in the caption of fig. S22.

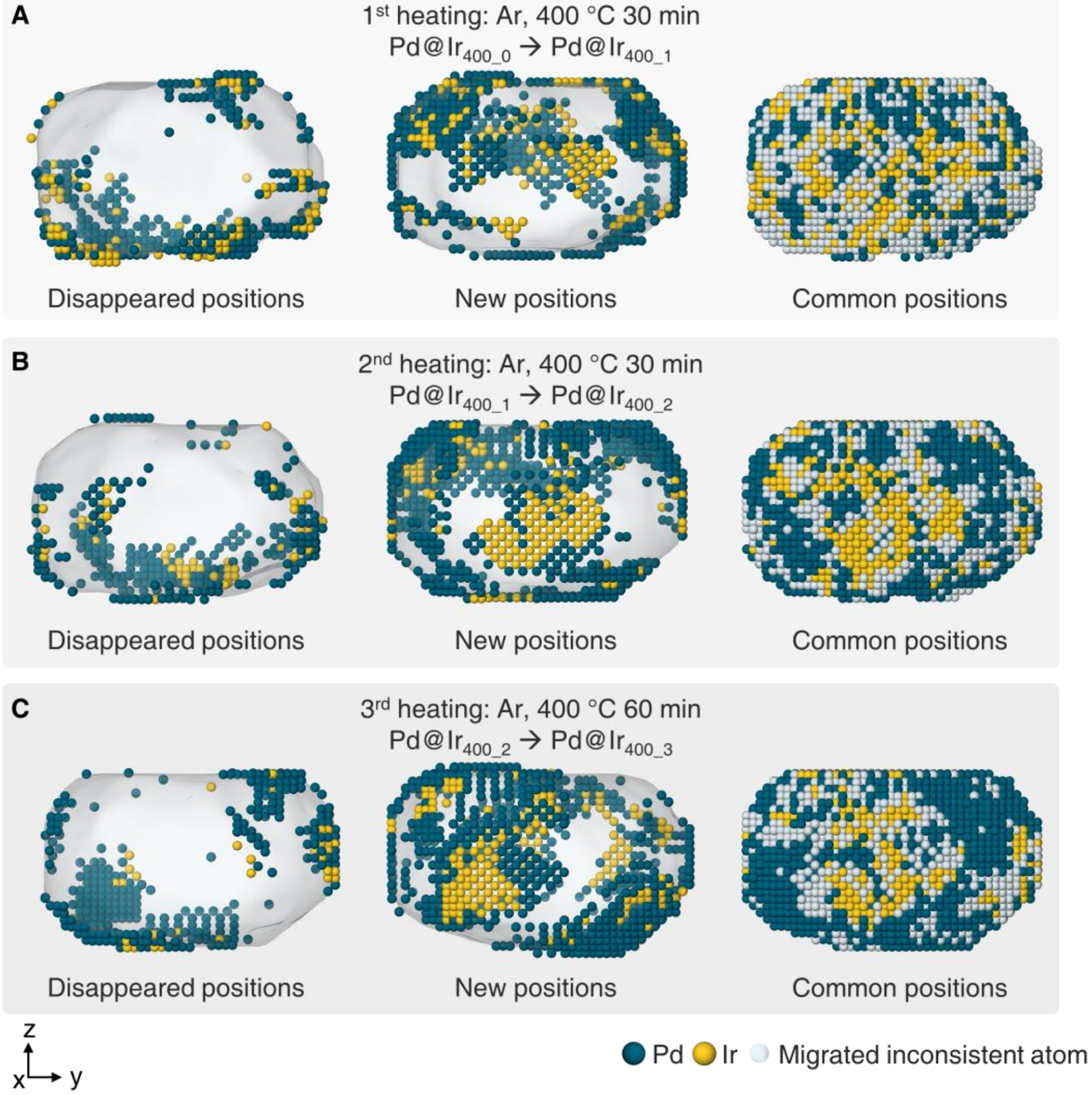

## Fig. S24. 4D evolution of atomic positions and chemical species in Particle-3.

(**A**) 3D renderings of Particle-3 before and after the 1st annealing (30 min annealing in Ar). (**B**) 3D renderings of Particle-3 before and after the 2nd annealing (60 min annealing in Ar). (**C**) 3D renderings of Particle-3 before and after the 3rd annealing (120 min annealing in Ar for two times).

The definitions of disappeared positions, new positions, common positions, migrated-inconsistent atoms, and migrated atoms are provided in the caption of fig. S22.

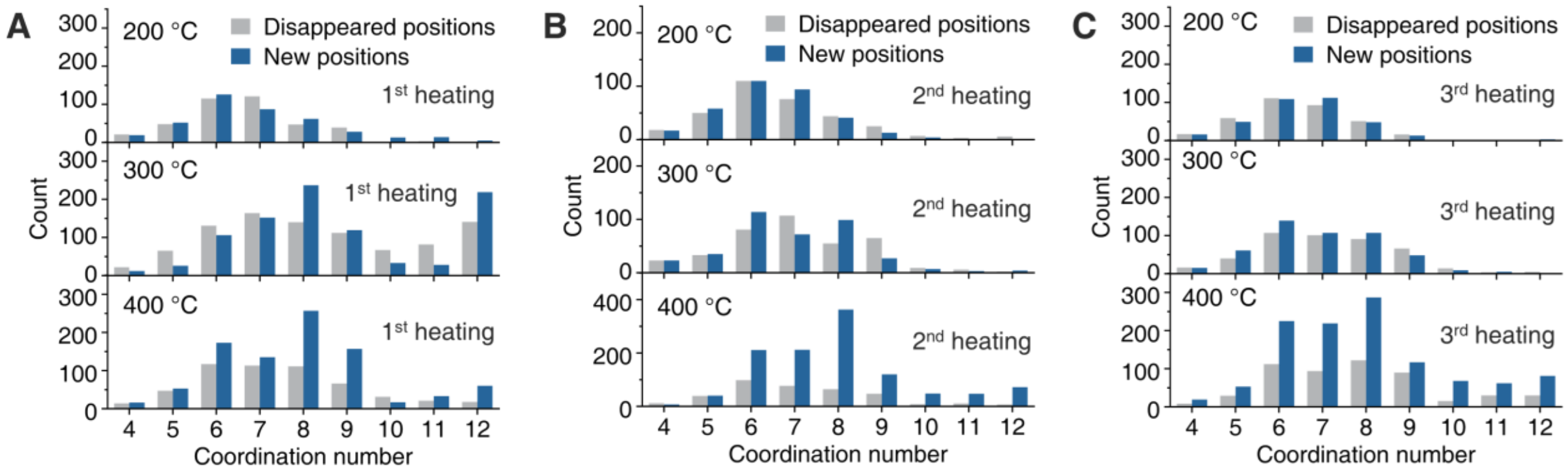

**Fig. 25. Changes in coordination numbers (CNs) of disappeared and new positions for Particles 1-3.**

(**A**) 1st annealing. (**B**) 2nd annealing. (**C**) 3rd annealing. Positions identified in previous stage but absent in later stage, are named as disappeared positions; positions newly identified in later stage are named as new positions.

In the 1st annealing: unpaired positions at 200 °C are primarily composed of atoms with CN=6–7, with structural changes dominated by surface reconstruction. At 300 °C and 400 °C, unpaired positions with CN = 12 emerged, indicating that subsurface began to reconstruct at these temperatures. The number of new positions at 400 °C far exceeds that of disappeared positions, pointing to the occurrence of Oswald ripening (fig. S20). Notably, more unpaired positions with CN = 12 were observed at 300 °C than at 400 °C, this is a difference attributable to variations between particles: pristine Particle-2 contains more concave and convex regions (a flatter distribution in fig. S26B than that in fig. S26C), and hence undergoes more pronounced changes during the 1st annealing (fig. S23).

In the 2nd and 3rd annealing: surface reconstruction at 200 °C remained similar to that in the 1st annealing, and still dominated by low-coordinated atoms (CN = 6–7). At 300 °C, structural changes become similar to that observed at 200 °C, primarily involving atoms with CN=6–7. This "milder" surface reconstruction compared to 1st annealing at 300 °C, is because the energetically unfavored atoms in concave and convex regions were flattened during 1st annealing (Fig. 1B). At 400 °C, Oswald ripening is also observed, and the structural evolution continued to involve atoms with CN=12, indicating that subsurface atoms continue to participate in the atomic diffusion.

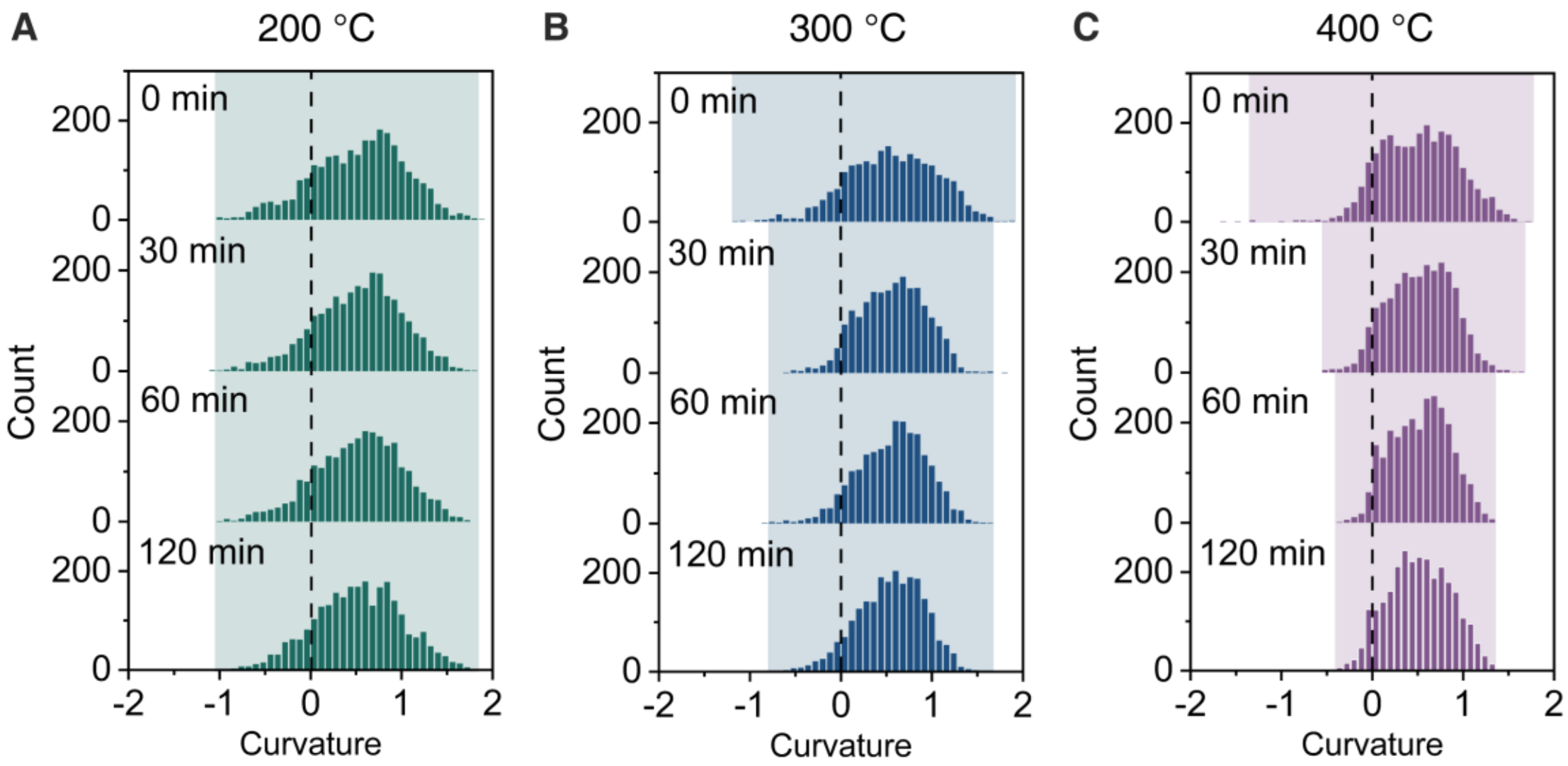

**Fig. S26. Curvature analysis for Particles 1-3.**

(**A**) Particle-1**.** (**B**) Particle-2**.** (**C**) Particle-3.

The pristine Particle-2 (B) exhibits a flatter curvature histogram compared to that of Particle-3 (C), indicating there are a greater number of concave and convex regions on Particle-2's surface.

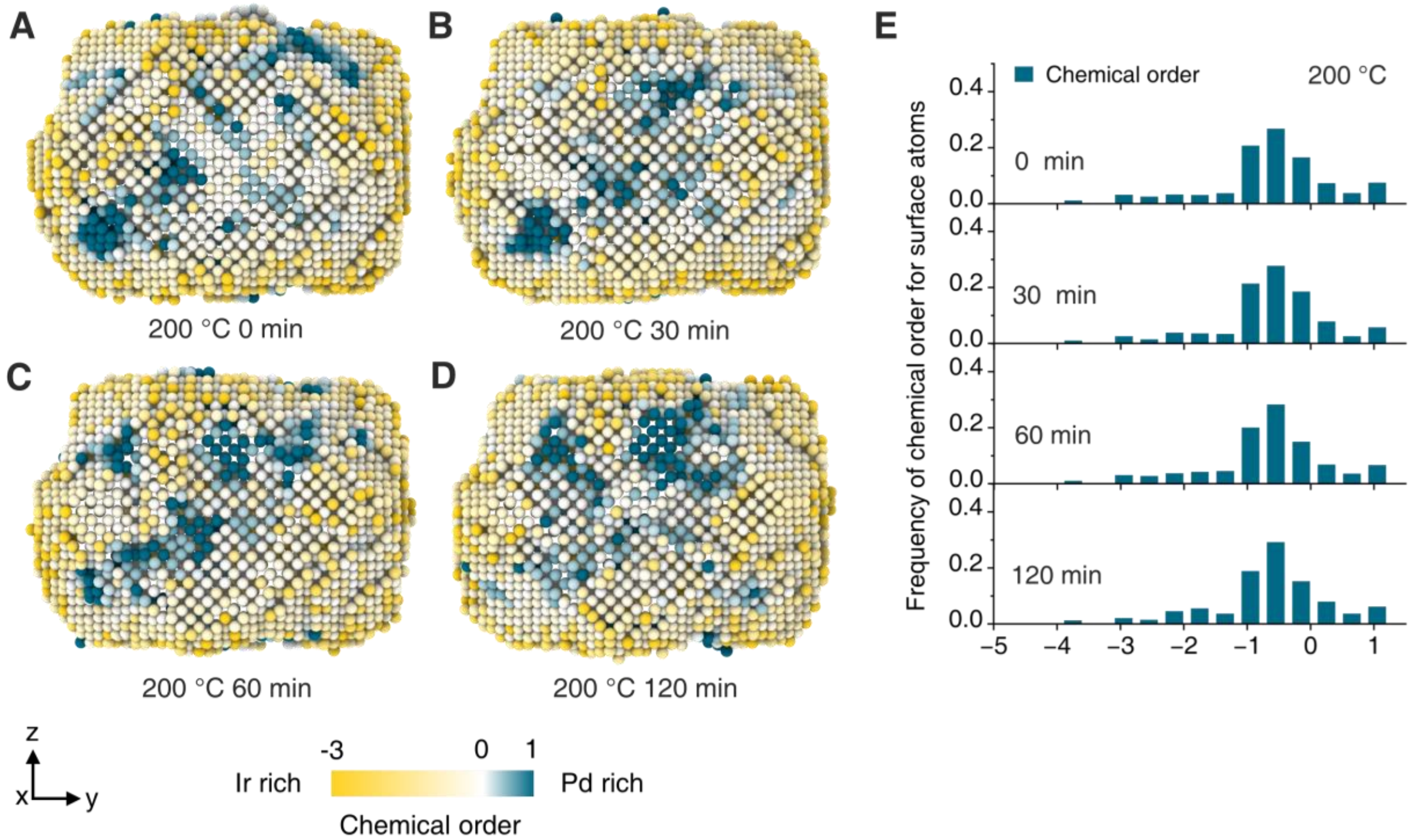

## Fig. S27. 3D renderings and histograms of CSROPs in Particle-1.

(**A** to **D**) 3D renderings of chemical order in Particle-1 after annealing for 0-min (A), 30-min (B), 60-min (C) and 120-min (D), respectively. The green, yellow and white colormap represents the Pd-rich regions, the Ir-rich regions and regions where the local composition equals the averaged Pd concentration in the whole particle, respectively. The color scale is normalized from -3 to 1. (**E**) Histograms of CSROPs for Particle-1.

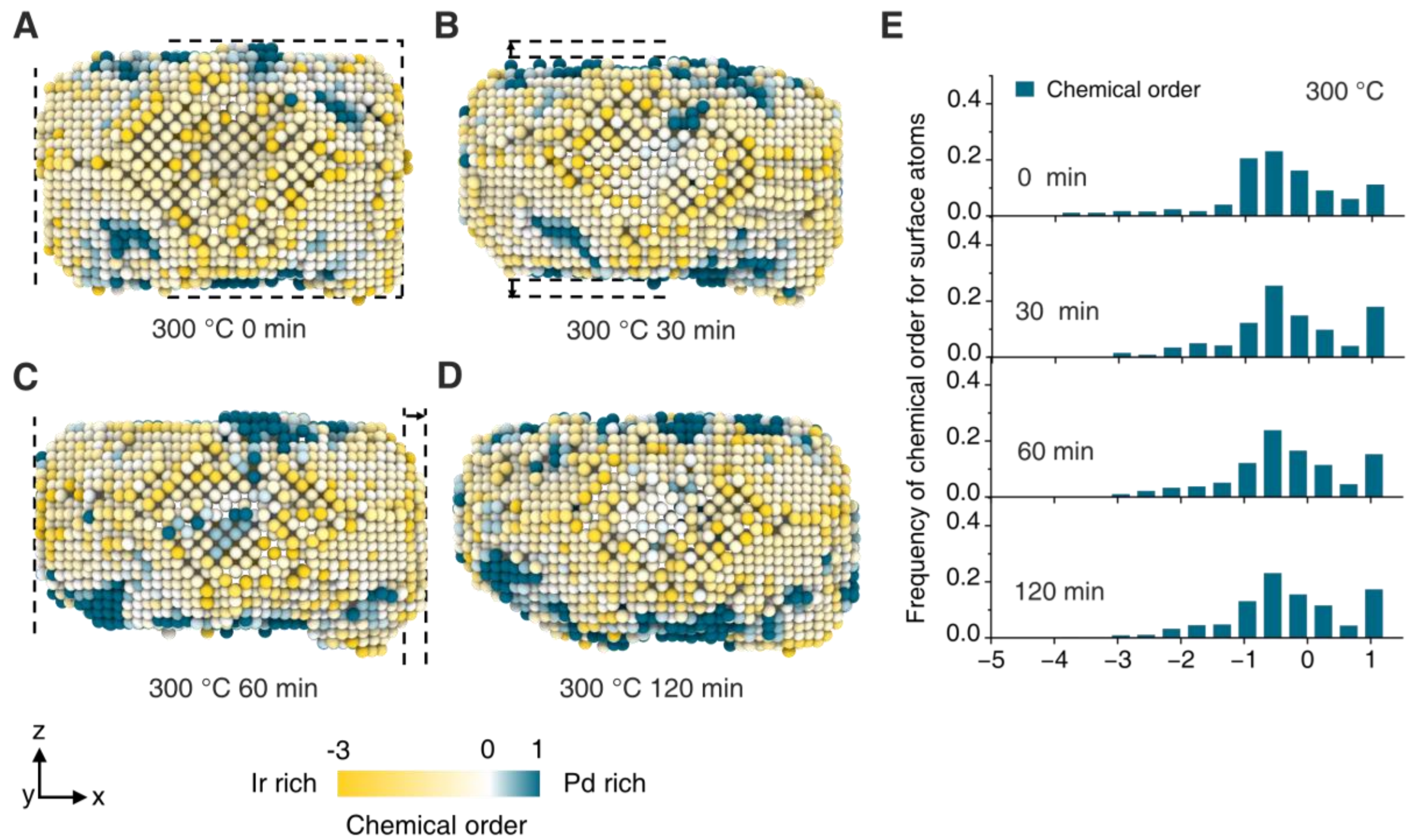

**Fig. S28. 3D renderings and histograms of CSROPs in Particle-2.**

(**A** to **D**) 3D renderings of chemical order in Particle-2 after annealing for 0-min (A), 30-min (B), 60-min (C) and 120-min (D), respectively. The green, yellow and white colormap represents the Pd-rich regions, the Ir-rich regions and regions where the local composition equals the averaged Pd concentration in the whole particle, respectively. The color scale is normalized from -3 to 1. (**E**) Histograms of CSROPs for Particle-2.

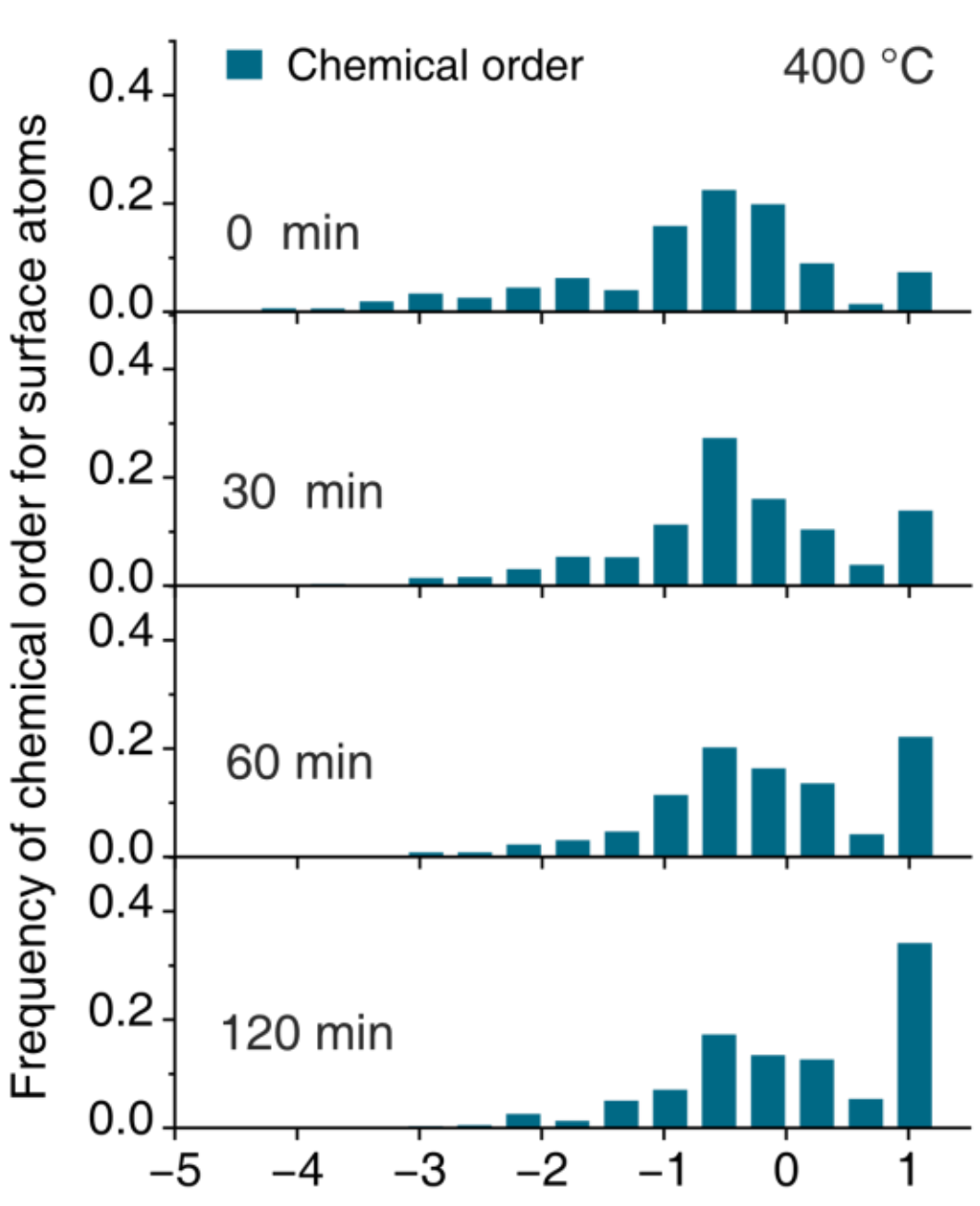

**Fig. S29. Histograms of CSROPs for Particle-3.**

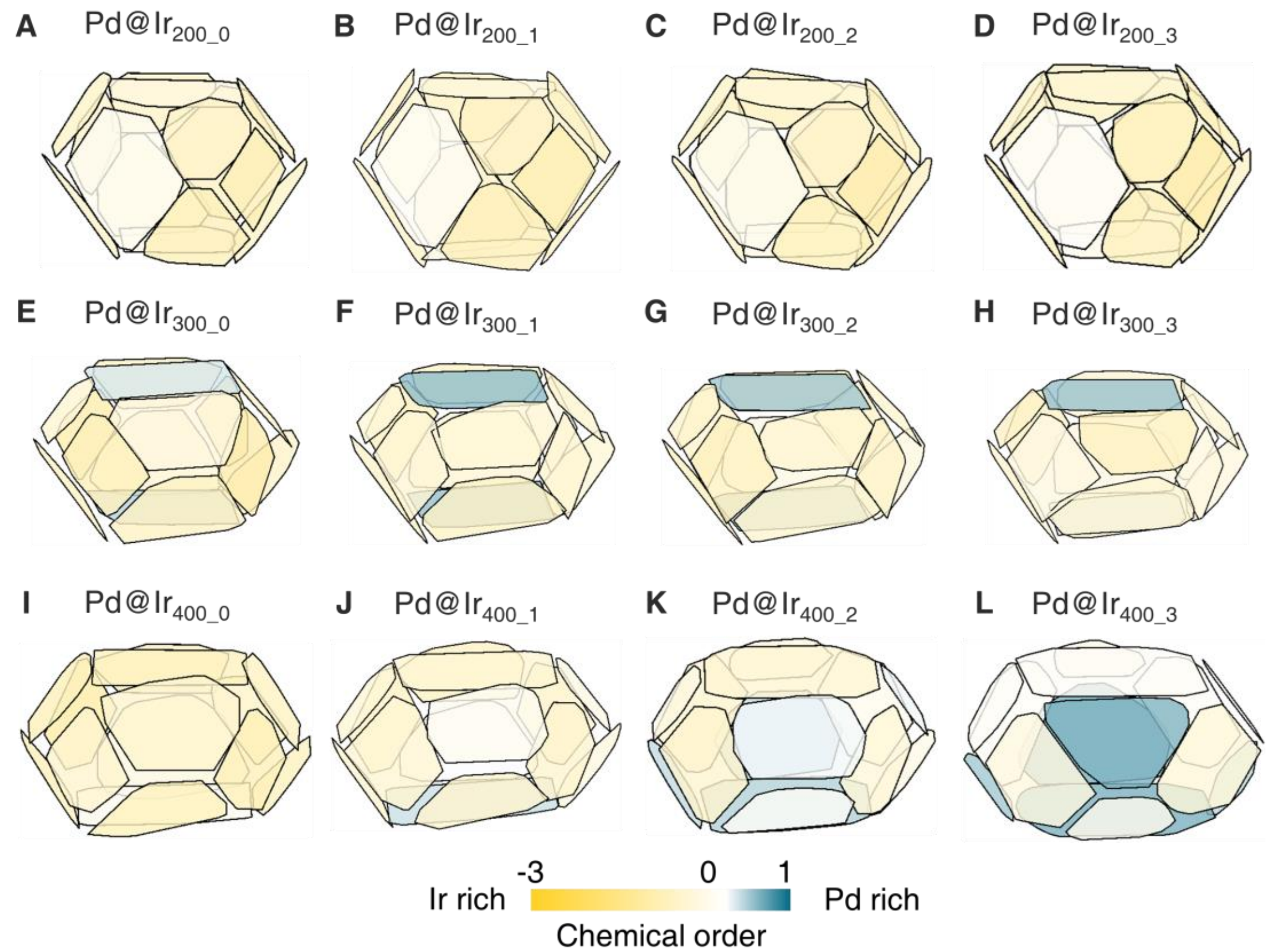

**Fig. S30. Renderings of averaged CSROPs in different orientations for Particles 1-3.**

(**A** to **D**) Averaged CSROPs of Particle-1 (A to D), Particle-2 (E to H) and Particle-3 (I to L), respectively. Surface atoms are grouped into eight {111}-oriented and six {100}-oriented regions, and the CSROPs are averaged within each region. The green, yellow and white colormaps represent the Pd-rich regions, the Ir-rich regions and the regions where the local composition equals the averaged Pd concentration in the whole particle, respectively. The color scale is normalized from -3 to 1.

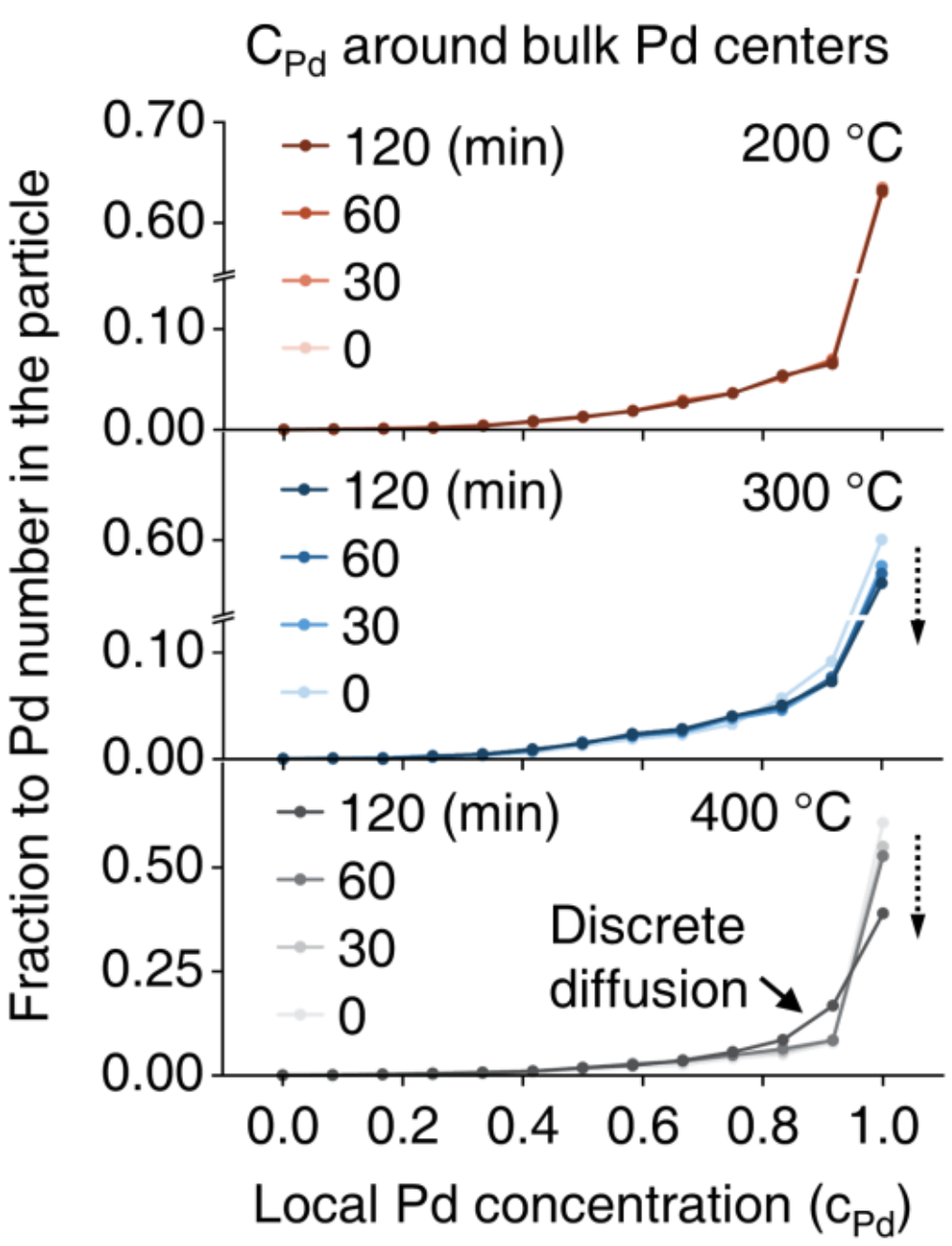

**Fig. S31. Overall distribution of local Pd concentrations ($C_{Pd}$) around bulk Pd centers.**

Bulk centers are defined as atoms with CN=12. $C_{Pd}$ (x axis) represents the Pd atomic fractions, Pd/(Pd+Ir), within the first coordination shells, indicating the local probability of finding Pd around a central atom. Dashed arrows highlight the decrease in the population of bulk Pd-rich centers.

At 200 °C, the $C_{Pd}$ distribution are nearly overlapped across all datasets. At 300 and 400°C, 1$^{st}$ annealing drives the counts of bulk Pd centers to decrease overall, indicating the surface Pd concentration increases. At 400 °C, the counts for $C_{Pd}$ = 83.3% and 91.7% increase at 120-min annealing, corroborated with the observation of discrete Ir diffusion in Fig. 3D.

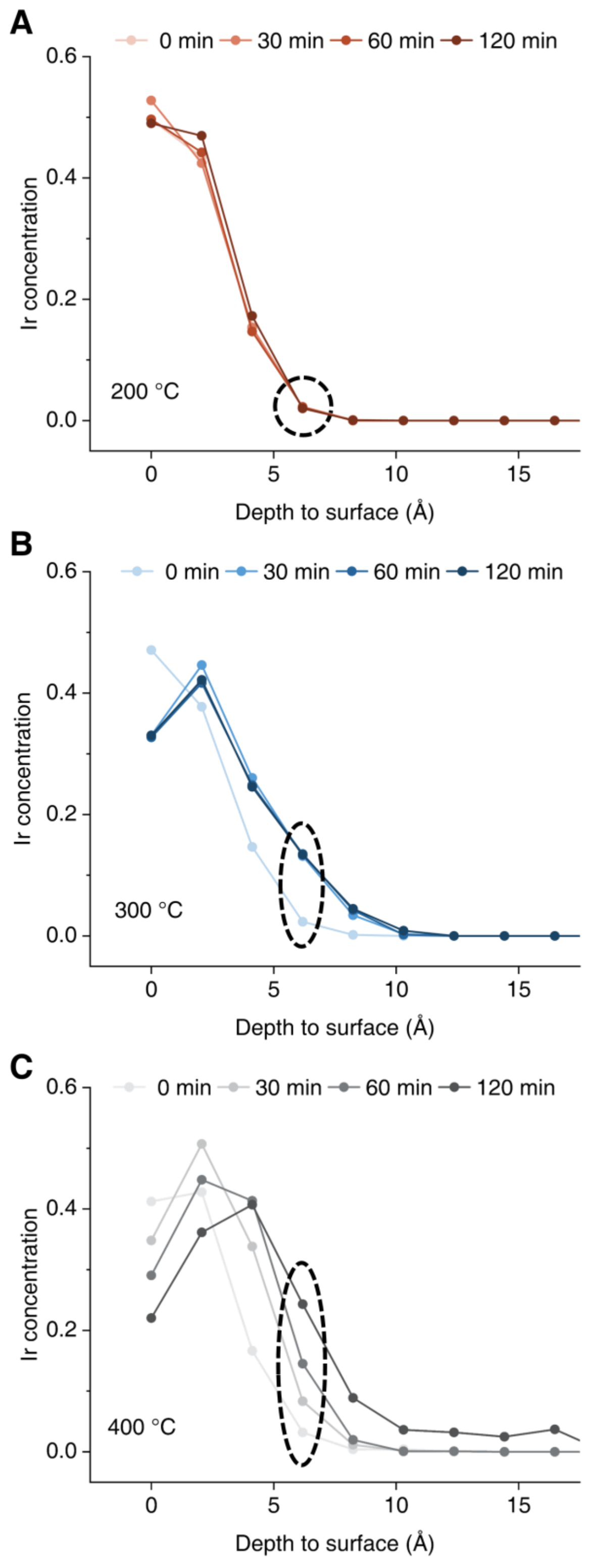

**Fig. S32. Ir radial concentrations at different depths in Particles 1-3.**
(**A** to **C**) Pd@Ir NPs for Particle-1 (a), Particle-2 (b), Particle-3 (c) at different annealing stages were "peeled" like an onion from the outside to get the atomic compositions in different depths.

**Table S1. AET data and analysis for Particle-1**

| Particle name | Pd@$Ir_{200_0}$ | Pd@$Ir_{200_1}$ | Pd@$Ir_{200_2}$ | Pd@$Ir_{200_3}$ |
|---|---|---|---|---|
| **Data collection*** | | | | |
| Tilt range (°) | -73°<br>72.5° | -73°<br>72.5° | -73°<br>72.5° | -73°<br>72.5° |
| Electron dose ($10^5$ $e^-$ $Å^{-2}$) | 3.0 | 3.0 | 3.0 | 3.0 |
| **Refinement†** | | | | |
| R (%)‡ | 6.4 | 6.5 | 6.5 | 6.5 |
| **Statistics of atoms** | | | | |
| Total number of Pd | 10639 | 10597 | 10641 | 10578 |
| Total number of Ir | 2852 | 2904 | 2856 | 2924 |
| **Common positions** | | | | |
| Number of positions | 13093 | 13093/13160§ | 13160/13148¶ | 13148 |
| Ratio to total (%) | 97.1 | 97.0/97.5§ | 97.5/97.4¶ | 97.4 |
| **Consistent atoms** | | | | |
| Number of Pd | 9766 | 9766/9818# | 9818/9773** | 9773 |
| Number of Ir | 2317 | 2317/2337# | 2337/2317** | 2317 |
| Ratio to total (%) | 92.3 | 92.3/92.4# | 92.4/92.0** | 92.0 |

* Operation parameters of datasets are: 300 kV voltage, 30 mrad convergence semi-angle, 40.6 mrad HAADF detector inner semi-angle and 200 mrad HAADF detector outer semi-angle, the pixel size is 0.352 Å.

† Real Space Iterative Reconstruction (RESIRE) algorithm (*44*) was used to reconstruct each experimental tilt series, with oversampling ratio being 4 and number of iterations being 300.

‡ a R-factor is used to calculate the difference between calculated and measured projections (*44*).

§ 13093 and 97.0% are the common atoms and ratio to total atoms between Pd@$Ir_{200_0}$ and Pd@$Ir_{200_1}$; and 13160 and 97.5% are the common atoms and ratio to total atoms between Pd@$Ir_{200_1}$ and Pd@$Ir_{200_2}$, respectively.

¶ 13160 and 97.5% are the common atoms and ratio to total atoms between Pd@$Ir_{200_1}$ and Pd@$Ir_{200_2}$; and 13148 and 97.4% are the common atoms and ratio to total atoms between Pd@$Ir_{200_2}$ and Pd@$Ir_{200_3}$, respectively.

\# 9766, 2317 and 92.3% are the consistent Pd and Ir atoms and ratio to common positions between Pd@$Ir_{200_0}$ and Pd@$Ir_{200_1}$; and 9818, 2337 and 92.4% are the consistent atoms and ratio to common positions between Pd@$Ir_{200_1}$ and Pd@$Ir_{200_2}$, respectively.

** 9818, 2337 and 92.4% are the consistent Pd and Ir atoms and ratio to common positions between Pd@$Ir_{200_1}$ and Pd@$Ir_{200_2}$; and 9773, 2317 and 92.0% are the consistent atoms and ratio to common positions between Pd@$Ir_{200_2}$ and Pd@$Ir_{200_3}$, respectively.

**Table S2. AET data and analysis for Particle-2**

| Particle name | $Pd@Ir_{300_0}$ | $Pd@Ir_{300_1}$ | $Pd@Ir_{300_2}$ | $Pd@Ir_{300_3}$ |
|---|---|---|---|---|
| **Data collection*** | | | | |
| Tilt range (°) | -75°<br>75° | -75°<br>75° | 75°<br>75° | 75°<br>75° |
| Electron dose ($10^5$ $e^-$ $Å^{-2}$) | 3.0 | 3.0 | 3.0 | 3.0 |
| **Refinement†** | | | | |
| R (%)‡ | 6.1 | 6.4 | 6.7 | 6.7 |
| **Statistics of atoms** | | | | |
| Number of Pd | 9130 | 9003 | 9071 | 9046 |
| Number of Ir | 2321 | 2456 | 2391 | 2466 |
| **Common positions** | | | | |
| Number of positions | 10527 | 10527/11078§ | 11078/11018¶ | 11018 |
| Ratio to total (%) | 91.9 | 91.9/96.7§ | 96.7/96.1¶ | 95.7 |
| **Consistent atoms** | | | | |
| Number of Pd | 8143 | 8143/8178** | 8178/8108†† | 8108 |
| Number of Ir | 1915 | 1915/1878** | 1878/1870†† | 1870 |
| Ratio to total (%) | 95.6# | 95.6#/90.8** | 90.8/90.6†† | 90.6 |

* Operation parameters of datasets are: 300 kV voltage, 30 mrad convergence semi-angle, 40.6 mrad HAADF detector inner semi-angle and 200 mrad HAADF detector outer semi-angle, the pixel size is 0.352 Å.

† Real Space Iterative Reconstruction (RESIRE) algorithm (*44*) was used to reconstruct each experimental tilt series, with oversampling ratio being 4 and number of iterations being 300.

‡ A R-factor is used to calculate the difference between calculated and measured projections (*44*).

§ 10527 and 91.9% are the common atoms and ratio to total atoms between $Pd@Ir_{300_0}$ and $Pd@Ir_{300_1}$; and 11078 and 96.7% are the common atoms and ratio to total atoms between Pd@Ir300_1 and $Pd@Ir_{300_2}$, respectively.

¶ 11078 and 96.7% are the common atoms and ratio to total atoms between $Pd@Ir_{300_1}$ and $Pd@Ir_{300_2}$; and 11018 and 96.1% are the common atoms and ratio to total atoms between $Pd@Ir_{300_2}$ and $Pd@Ir_{300_3}$, respectively.

\# Surface reconstruction is more active here for the existing of more low-coordinated atoms in Particle-2. This high chemical consistency is attributed to the low ratio of common positions, where more atoms leave their original positions and lower the common positions' number (fig. S23A).

** 8143, 1915 and 95.6% are the consistent Pd and Ir atoms and ratio to common positions between $Pd@Ir_{300_0}$ and $Pd@Ir_{300_1}$; and 8178, 1878 and 90.8% are the consistent atoms and ratio to common positions between $Pd@Ir_{300_1}$ and $Pd@Ir_{300_2}$, respectively.

†† 8178, 1878 and 90.8% are the consistent Pd and Ir atoms and ratio to common positions between $Pd@Ir_{300_1}$ and $Pd@Ir_{300_2}$; and 8108, 1870 and 90.6% are the consistent atoms and ratio to common positions between $Pd@Ir_{300_2}$ and $Pd@Ir_{300_3}$, respectively.

**Table S3. AET data and analysis for Particle-3**

| Particle name | Pd@Ir$_{400_0}$ | Pd@Ir$_{400_1}$ | Pd@Ir$_{400_2}$ | Pd@Ir$_{400_3}$ |
|---|---|---|---|---|
| **Data collection*** | | | | |
| Tilt range (°) | -73°<br>73° | -73°<br>72.5° | -73°<br>72.5° | -73°<br>70° |
| Electron dose ($10^5$ $e^-$ Å$^{-2}$) | 2.6 | 2.6 | 2.6 | 2.6 |
| **Refinement†** | | | | |
| R (%)‡ | 6.6 | 5.5 | 5.5 | 5.7 |
| **Statistics of atoms** | | | | |
| Number of Pd | 12035 | 11906 | 12532 | 12819 |
| Number of Ir | 2797 | 3283 | 3415 | 3733 |
| **Common positions** | | | | |
| Number of positions | 14286 | 14286/14286§ | 14286/15416¶ | 15416 |
| Ratio to total (%) | 96.3 | 94.0/94.0§ | 89.6/96.7¶ | 93.1 |
| **Consistent atoms** | | | | |
| Number of Pd | 10309 | 10309/10672# | 10672/10661** | 10661 |
| Number of Ir | 1717 | 1717/2227# | 2227/2034** | 2034 |
| Ratio to total (%) | 84.2 | 84.2/90.3# | 90.3/82.4** | 82.4 |

* Operation parameters of datasets are: 300 kV voltage, 30 mrad convergence semi-angle, 40.6 mrad HAADF detector inner semi-angle and 200 mrad HAADF detector outer semi-angle, the pixel size is 0.352 Å.

† Real Space Iterative Reconstruction (RESIRE) algorithm (*44*) was used to reconstruct each experimental tilt series, with oversampling ratio being 4 and number of iterations being 300.

‡ a R-factor is used to calculate the difference between calculated and measured projections (*44*).

§ 14286 and 94.0% on the left are the common atoms and ratio to total atoms between Pd@Ir$_{400_0}$ and Pd@Ir$_{400_1}$; and 14286 and 94.0% on the right are the common atoms and ratio to total atoms between Pd@Ir$_{400_1}$ and Pd@Ir$_{400_2}$, respectively.

¶ 14286 and 89.6% are the common atoms and ratio to total atoms between Pd@Ir$_{400_1}$ and Pd@Ir$_{400_2}$; and 15416 and 96.7% are the common atoms and ratio to total atoms between Pd@Ir$_{400_2}$ and Pd@Ir$_{400_3}$, respectively.

# 10309, 1717 and 84.2% are the consistent Pd and Ir atoms and ratio to total atoms between Pd@Ir$_{400_0}$ and Pd@Ir$_{400_1}$; and 10672, 2227 and 90.3% are the consistent atoms and ratio to common positions between Pd@Ir$_{400_1}$ and Pd@Ir$_{400_2}$, respectively.

** 10672, 2227 and 90.3% are the consistent Pd and Ir atoms and ratio to total atoms between Pd@Ir$_{400_1}$ and Pd@Ir$_{400_2}$; and 10661, 2034 and 82.4% are the consistent atoms and ratio to common positions between Pd@Ir$_{400_2}$ and Pd@Ir$_{400_3}$, respectively.

**Movie S1**

Raw ADF-STEM images recorded at 900°C. One second in the movie represents approximately 2.1 s in the heating experiment. Scale bar is 5 nm for the field of view in the movie.